%% file: main.tex
\documentclass{article}
\usepackage{hyperref}
\usepackage{authblk}
\usepackage{fix-cm}
\usepackage{graphicx} 
\usepackage[a4paper, margin=1in]{geometry}
\usepackage{tikz}
\usetikzlibrary{arrows.meta, positioning, shapes, calc, math}
\usepackage{float}
\usepackage[caption=false]{subfig}
\usepackage{algorithm}
\usepackage{algpseudocode}
\usepackage{amsmath}
\usepackage{amssymb}
\usepackage{array}
\usepackage{url}
\usepackage{pifont}

\newcommand{\cmark}{\ding{51}}%
\newcommand{\xmark}{\ding{55}}%

\begin{document}

\title{A Full-Stack Platform Architecture \\for Self-Organised Social Coordination}


\author[1,2]{Matthew Scott}
\author[1,3]{Jeremy Pitt}
\affil[1]{Imperial College London}
\affil[2,3]{\textit {\{matthew.scott18, j.pitt\}@imperial.ac.uk}}

\maketitle

\begin{abstract}
    To mitigate the restrictive centralising and monopolistic tendencies of \textit{platformisation}, we aim to empower local communities by democratising platforms for self-organised social coordination. Our approach is to develop an open-source, full-stack architecture for platform development that supports ease of distribution and cloning, generativity, and a variety of hosting options. The architecture consists of a meta-platform that is used to instantiate a base platform with supporting libraries for generic functions, and plugins (intended to be supplied by third parties) for customisation of application-specification functionality for self-organised social coordination. Associated developer- and user-oriented toolchains support the instantiation and customisation of a platform in a two-stage process. This is demonstrated through the proof-of-concept implementation of two case studies: a platform for regular sporting association, and a platform for collective group study. We conclude by arguing that self-organisation at the application layer can be achieved by the specific supporting functionality of a full-stack architecture with complimentary
    developer and user toolchains.
\end{abstract}


\section{Introduction}

Platformisation is the process describing the emergence and proliferation of platforms \cite{poell2019platformisation}, where a `platform' is an amorphous concept referencing any software infrastructure that promotes interaction in any sector of the \textit{Digital Society}, e.g., e-commerce, social media, and productivity apps \cite{srnicek2017platform}.


Alongside significant benefits, network effects at the application level have also created monolithic and monopolistic platforms, often owned by private and
commercial organisations. This restricts users to only using functions supported by these systems, and opens up possibilities for misuse: e.g., surveillance capitalism \cite{zuboff2023age}, 
AI-driven concentration of power \cite{pittArchReemp}, and algorithmic misinformation \cite{algorithmicMisinfo}.

Alternatively, local communities can be re-empowered by democratising platforms for social coordination. Through the provision of an open-source,
full-stack architecture for platform development, we can support ease of distribution and cloning, generativity, and a variety hosting options.

Ease of distribution
and cloning reduces barriers to entry, as one community can take advantage of solutions already developed an deployed by another \cite{NowakBubbleTheory}. 
Furthermore, it resists opportunities to buy up or buy out the community.
This enables a nascent community to co-opt, adopt, and adapt an existing and solution
rather than
trying to construct one from scratch. Generativity \cite{zittrain}, in relation to a tool, is the feature that enables that tool to be used in the innovation of other tools, that were not imagined by the original tool-maker. A variety of hosting options facilitates server-side transparency, offering communities multiple choices for operating their coordination systems according to their own resources and preferences. 

This paper presents one such full-stack architecture, called
\textit{PlatformOcean}. This consists of a meta-platform that is used to instantiate a base platform with supporting libraries for generic functions, and plugins (intended to be supplied by third parties) for customisation of application-specification functionality for self-organised social coordination. Associated developer- and user-oriented toolchains support the instantiation and customisation of a platform in a two-stage process. 
It is available open-source at \url{github.com/MattSScott/PlatformOcean}.

The critical supra-functional requirement of this architectural approach is to
support self-organised social coordination. This is demonstrated through the implementation of two proof-of-concept case studies using both the same and different plugins: a platform for regular sporting association, and a platform for collective group study.

Accordingly, this paper is structured as follows.
Section~\ref{sec:FSA} presents the design and implementation of the 
full-stack architecture: the meta-platform, platform, and plugins.
Section~\ref{sec:toolchains} describes the developer and user toolchains
which support the process of platform instantiation and application development.
Section~\ref{sec:caseStudies} analyses the two case studies. 
After a review of related and further work in Section~\ref{sec:furtherWork},
we summarise and conclude in Section~\ref{sec:sumConc}, 
arguing that self-organisation at the application layer can be achieved by the specific supporting functionality of a full-stack architecture with complimentary developer and user toolchains.



\section{Full-Stack Architecture}\label{sec:FSA}

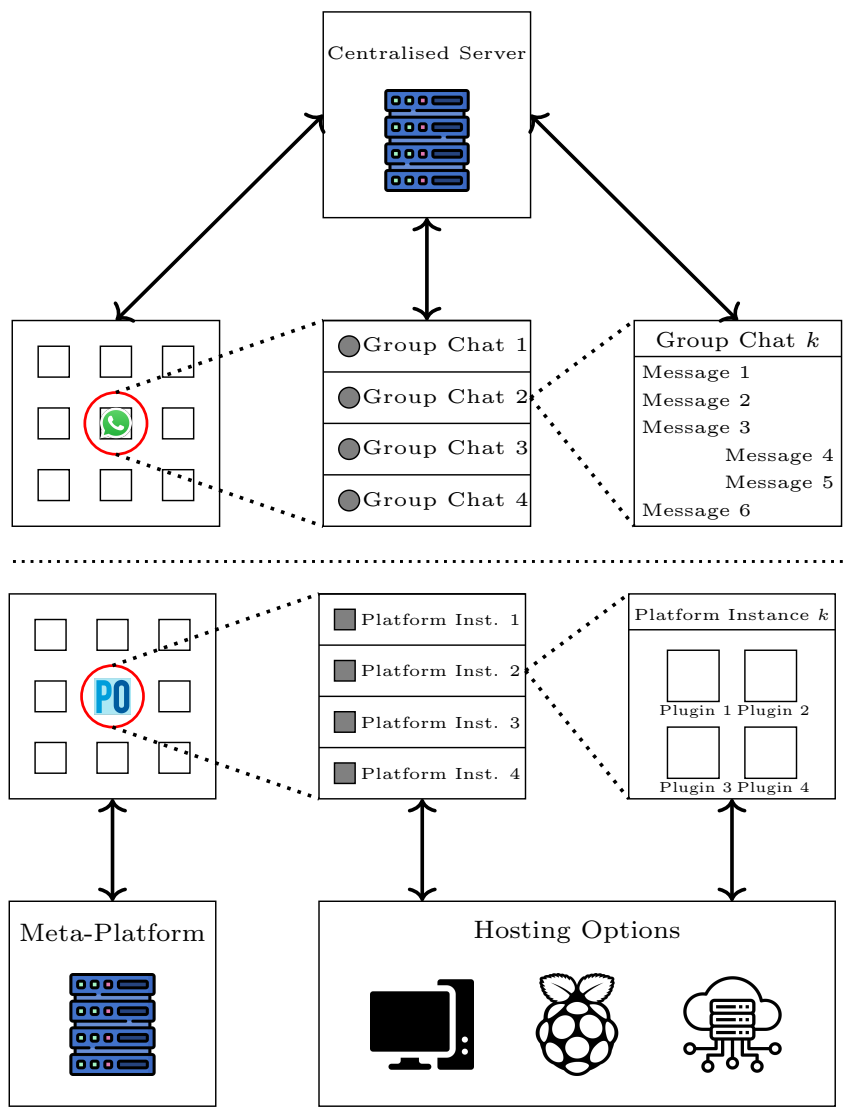
\begin{figure}[htb]
\centering
\resizebox{0.7\linewidth}{!}{\input{tikz_plots/p1.tex}}
\resizebox{0.7\linewidth}{!}{\input{tikz_plots/p2.tex}}
\\[1em]
\resizebox{0.7\linewidth}{!}{\input{tikz_plots/p3.tex}}
\caption{Full-stack system architecture for platform and meta-platform, compared and contrasted with \textit{WhatsApp}'s architecture}\label{fig:poVsWhatsapp}
\end{figure}

At a high-level of abstraction, the \textit{PlatformOcean} full-stack architecture shares many similarities with conventional social-coordination system architectures. Taking \textit{WhatsApp} for example (i.e., a centralised,  multimedia chat app), the `platform' can be thought of as the app itself. Continuing this analogy, the concept of a meta-platform as a means of instantiating the platform can be compared to the \textit{app store}. Furthermore, in the way that one user can be a member of multiple group chats simultaneously, so too can they be a member of multiple \textit{PlatformOcean} instances. A visual comparison is illustrated in Figure~\ref{fig:poVsWhatsapp}.

\vspace*{-0.75em}

These approaches diverge, though, in the degree of centralisation. This is highlighted in Figure~\ref{fig:poVsWhatsapp} by inverting the position of the server in the \textit{WhatsApp} and \textit{PlatformOcean} stacks. In the \textit{WhatsApp} architecture, the central component is a server handling instantiation of the app and the distributed message passing; all messages sent over the app are relayed (and stored) on a centralised server owned by the company. In contrast, in the \textit{PlatformOcean} architecture, the only role of the centralised server is to provide a single point of contact for instantiating the server, client app, and plugins. Subsequently, all data is stored on, and all messages are relayed through, a localised platform server with multiple hosting options: for example a desktop, \textit{raspberry pi} or on the cloud.

Reifying this abstraction, this section decomposes the full-stack \textit{PlatformOcean} architecture into three stages. Firstly, at the top level, Section~\ref{sec:metaPlat} defines the \textit{meta-platform} (i.e., a platform for platforms) which serves as a single, centralised point of contact for instantiating the \textit{platform} instances. Secondly, Section~\ref{sec:platform} defines these platform instances in their `vanilla' state, and how they provide the necessary tools for user registration, data storage, and instance discovery. Finally, in order to provide the bespoke platform functionality, Section~\ref{sec:plugins} describes how these platforms can be customised with \textit{plugins} -- dynamically injected software components that provide a user interface (UI) similar to a ``micro-app'' or ``micro frontend'' \cite{zateishchikov2023scaling}, allowing for users to interact via datagram messages sent across the platform.


\subsection{Meta-Platform}\label{sec:metaPlat}

The \textit{meta-platform} provides a single point of access for both users and developers. For a developer, the meta-platform acts as plugin repository -- i.e., an endpoint for plugin developers to submit their designs for subsequent download/inclusion in a platform instance. Alg.~\ref{alg:pluginReg}, line 9 formalises this process.

\begin{algorithm}[htb]
\caption{PlatformOcean Plugin Registry Algorithm}
\label{alg:pluginReg}
    \begin{algorithmic}[1]
        \State \textbf{define} \textit{bundleable} $\supseteq \{\textit{designFiles}, \textit{styleFiles}, \textit{imports}\}$\\

        \Procedure{\textit{bundle}}{\textit{bundleable}}
        \State \textbf{generate} \textit{remoteEntry} \textbf{from} \textit{bundleable}
        \State $\textit{injectable} \gets \textit{bundleable} \cup \textit{remoteEntry}$
            \State \Return \textbf{hash}(\textbf{minify}(\textbf{transpile}(\textit{injectable}))) 
        \EndProcedure \\

        \Procedure{\textit{acceptPlugin}}{\textit{metaPlatform, injectable}}
            \State \textbf{Local variable}:
            \State \quad struct \textit{plugin p}:
            \State \quad\quad \textit{pluginNode}: \textit{injectable}
            \State \quad\quad \textit{pluginKey}: \textit{UUID}
            \State \quad\quad \textit{pluginLocation}: $\textit{string}$
            \State $p \gets \textbf{new} \ \textit{plugin}(\textit{injectable}, \textit{UUID()}, \textbf{write loc})$
            \State \textit{persist}(\textit{metaPlatform}, $p$)
            \State \textbf{expose} \textbf{route} \textbf{to} \textit{p.pluginNode.remoteEntry}
        \EndProcedure \\

    \Procedure{\textit{persist}}{\textit{metaPlatform, plugin}}
        \State $\textit{key} \gets \textit{plugin.pluginKey}$
        \State \textbf{if} $\textit{key} \in \textit{metaPlatform}$ \textbf{then}
        \State \quad \quad \textit{versionHistory} $\gets \textit{map(key, versionHistory)}$
        \State \textbf{else}
        \State \quad \quad \textit{versionHistory} $\gets \{\}$
        \State \Return $\textit{versionHistory} \gets \textit{versionHistory} \cup \{\textit{plugin}\}$
    \EndProcedure
    \end{algorithmic}
\end{algorithm}

The \textbf{bundleable} (Alg.~\ref{alg:pluginReg}, line 1) is the key data structure underpinning \textit{PlatformOcean's} operation. In order to provide a fully customisable, generic system architecture, developers must be able to codify plugins unrestrictedly -- that is, without having to conform too tightly to a given specification. Therefore, developers can leverage external libraries, various stylings and multiple design files to specify their plugin UI. The sole requirement is that the plugin, when finished, can be \textit{bundled} (i.e., statically analysed and converted into one or more optimized output files) in order to be sent over a network and uploaded to the \textit{meta-platform}. Hence, a bundleable is defined (recursively) as anything that can be \textit{bundled}. In this context, a bundleable manifests itself as extending a set of multiple, individually bundleable components. The minimal set of requirements for the a bundleable is: \textit{design files} -- the raw JSX code that defines the UI of the plugin; \textit{style files} -- the raw CSS code that defines the stylings of the plugin component (in an HTML-compatible format); and the \textit{imports} -- the set of external libraries used in specifying the plugin.

\begin{figure}[htb]
\centering
\resizebox{\linewidth}{!}{\input{tikz_plots/pluginReg.tikz}}
\caption{Full-stack plugin registry event loop}\label{fig:messageDist}
\end{figure}
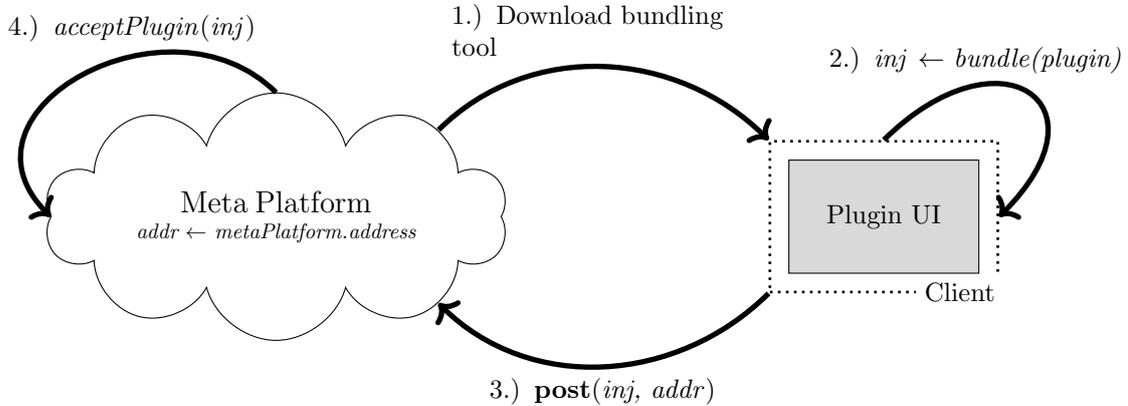

Having fully specified a plugin, a user bundles their bundleable into a platform-appropriate format by using the toolchain specified later in Section~\ref{sec:devTools}. This toolchain generates a \textit{remote entry} `meta' file (i.e., a file describing the file structure) that can be remotely accessed, allowing the plugin to be injected into a platform instance. Once the \textit{remote entry} file has been attached to the bundle, thereby creating an \textbf{injectable}, the injectable is \textit{transpiled} into a uniform JSX representation, \textit{minified} to reduce the size of the plugin bundle, and \textit{hashed} to give the encoded plugin injectable a unique identifier (thereby allowing for multiple plugins of the same name).

The role of the \textit{meta-platform}, then, is to run a protocol that accepts the uploaded plugin, persists it in a database, and exposes a route to the \textit{remote entry} file. This requires three additional components to exist on the \textit{meta-platform} server: a static endpoint for the upload route, a database schema for persistence, and a dynamic file store for hosting and exposing the address of the plugin's entry point.

The upload route simply becomes a static endpoint: \texttt{<ADDRESS>/upload}, say, which accepts a \textbf{POST} request with a \textbf{multipart file} payload. The plugin bundle can then be unzipped (as the bundle must be sent across the internet as a serialisable zip file), written to the server's local memory, and exposed as a static file using the paradigm of a static file server.  The plugin's \textit{metadata} (i.e., data about the data) persists, which helps with organisation and identification, as well as minimising the footprint of the response when a client queries the database for the list of available plugins. To do this, the \textit{meta-platform} stores a uniquely generated \textit{UUID} for each plugin, and (as strings) its name, the URL of its remote entry file, and the physical file location (on the server). A full event loop, from creation to persistence, is shown in Figure~\ref{fig:messageDist}.



\subsection{Platform} \label{sec:platform}

The platform can reductively be thought of as a data storage and distribution system. The backend of the platform, when instantiated as a locally-hosted server, stores users, datagrams, and (a reference to) the plugins used to customise it. The frontend, however, yields a user interface (UI) for interacting with the backend, through a set of predefined protocols passed to the abstract plugin components. Accordingly, we describe the \textit{backend} operation of the platform in Section~\ref{sec:backend}, and its \textit{frontend} in Section~\ref{sec:frontend}.

\subsubsection{Backend}\label{sec:backend}

The baseline purpose of the platform is twofold. At a `vanilla' level, the platform acts as a `beacon', broadcasting an identifying signal across a multicast address such that it can be used in the \textit{simple service discovery protocol} (SSDP). Furthermore, the platform is used to register users by generating an ID for them, and persisting their usernames and passwords. The protocol for joining a platform is specified in Alg.~\ref{alg:platJoin}, line 19. Here, the \textbf{client} data structure (i.e., the object used to characterise the user joining the platform) must track the address of the platform, the unique ID of the client, and the visual state of the client (e.g., the landing page for platform discovery or the homepage of a joined platform instance).

To access a prospective platform, a user can leverage the `beacon' operation of the platform, as well as the SSDP tool (described later in Section~\ref{sec:devTools}). This allows the user to access the address of all platforms on the local network. Having selected an address, the user can then either retrieve their unique platform \textit{clientID} if they're a member, or submit a request to join. Upon a successful retrieval of the \textit{clientID}, the client state will then update to render the homepage of the selected platform instance. 

\begin{algorithm}[htb]
\caption{PlatformOcean Instance Joining Algorithm}
\label{alg:platJoin}
\begin{algorithmic}[1]
\State \textbf{define} struct \textit{client}:
\State \quad \textit{client.platformAddress} $\gets \textbf{null}$
\State \quad \textit{client.clientID} $\gets \textbf{uuid.null}$
\State \quad \textit{client.state} $\gets *\textbf{landing page}*$ \\

\Procedure{\textit{discovery}}{\textit{network}}
\State \Return $\{ip \in network : \textbf{fetch}(ip) = *\textbf{passkey}*\}$
\EndProcedure \\

\Procedure{\textit{getUserID}}{\textit{endpoint, name, pass}}
\State \textbf{if} \textit{userExists(endpoint.names, endpoint.passes)} \textbf{then}
\State \quad \quad \textit{userHash} $\gets \textit{endpoint}.\textbf{lookup}(\textit{name, pass})$
\State \textbf{else}
\State \quad \quad \textit{endpoint}.\textbf{register}(\textit{name, pass})
\State \quad \quad \textit{userHash} $\gets$ \textbf{uuid.New()}

\State \Return \textit{userHash}
\EndProcedure \\

\Procedure{\textit{signOn}}{\textit{client, network}}
    \State \textit{endpoint} $\gets ip \in \textit{discovery}(network)$
    \State \textit{client.platformAddress} $\gets$ \textit{endpoint}
    \State \textit{name, pass} $\gets$ $*\textbf{user input}*$ 
    \State \textit{client.clientID} $\gets$ \textit{getUserID}(\textit{endpoint, name, pass})
    \State \textbf{if} \textit{client.state} = $*\textbf{landing page}*$
    \State \quad \textit{client.state} $\gets *\textbf{plugin access}*$
\EndProcedure
\end{algorithmic}
\end{algorithm}

When the platform is customised with plugins, it gains two further features. Firstly, the platform can be thought of as a `coordinating agent' for all of the connected clients in the network. It stores the current list of plugins that have been added to the platform, as well as a unique key to uniquely identify each plugin instance. This is different to the ID in the \textit{meta-platform's} plugin metadata, as the platform can host multiple copies/instances of a single type of plugins. Consequently, a single plugin identifier maps to multiple instantiations. Secondly, the platform stores all recorded plugin data (the datagrams specified by the plugin) and redistributes it across all of the connected clients via a websocket connection. This allows for real-time interactions over the platform's plugins. 

This message distribution algorithm is contingent on two datatypes: the \textbf{request} (representing an inbound message from the client to server) and the \textbf{response} (representing an outbound message from server to client). Common to these datatypes is the \textit{datagram} object in Table~\ref{tab:messageSchema}: all arbitrary datagrams must encode the provenance of the datagram through who it came from (its \textit{senderID}) and where it came from (its \textit{pluginID}), as well as the \textit{payload} of the datagram itself.

\begin{table}[htb]
    \centering
    \caption{Schema to fully characterise arbitrary datagram in platform}
    \label{tab:messageSchema}
    \begin{tabular}{c|c|c|c}
       \textbf{Parameter}  & \textbf{Range} & \textbf{Request} & \textbf{Response}  \\
       \hline
        \textit{datagramID} & \textit{UUID} & \xmark & \cmark \\
        \textit{senderID} & \textit{UUID} & \cmark & \cmark \\
        \textit{pluginID} & \textit{UUID} & \cmark & \cmark \\
        \textit{payload} & $\mathit{JsonNode} \supseteq \textbf{Object}$ & \cmark & \cmark\\
        \textit{protocol} & \textit{enum\{create, update, delete\}} & \xmark & \cmark \\
        \textit{shoudlPersist} & \textit{bool} & \cmark & \xmark
    \end{tabular}
\end{table}

A \textbf{request} from the client attaches a boolean \textit{shouldPersist} field to the datagram, to inform the platform whether or not to commit the message to memory. Upon receiving this data, the backend strips this field and generates a \textbf{response} schema. The backend generates and attaches a unique \textit{datagramID}, to identify the datagram, as well as the \textit{protocol} with which the datagram was sent -- i.e., \textit{create}, \textit{update}, or \textit{delete}. This allows the frontend to conditionally handle the received \textbf{response}.


\subsubsection{Frontend}\label{sec:frontend}

Having specified the representation of arbitrary datagrams, and the backend system that allows for their redistribution and storage, we now describe the operations available on the platform's fronted for creating and distributing these datagrams. 

The frontend component of the platform is responsible for providing a user interface for interaction with its backend. In serving as an implementation of a decentralised, plugin architecture, the platform frontend is capable of dynamically connecting to multiple backends, and supporting a wide array of grammars specified on various plugins. This therefore allows a single frontend tool to connect to multiple different platform instances with the toolchain described in Section~\ref{sec:userTools}, using the protocol described in Alg.~\ref{alg:platJoin}, line 19.

In order to support a variety of plugins -- all of which will be designed with different grammars and functionalities -- the frontend tool provides a \textit{wrapper} component. This \textit{plugin wrapper} provides a set of attributes and methods for interacting with an arbitrary platform instance, to comply with the protocol defined in Section~\ref{sec:backend}. This list of frontend methods and attributes is shown in Table~\ref{tab:pluginAtts}.

\begin{table}[htb]
    \caption{Attributes and methods provided by wrapper component}\label{tab:pluginAtts}
    \centering
    \begin{tabular}{c|c}
    \multicolumn{2}{c}{\centering \textbf{Attributes}}\\
    \hline
    \textbf{Name} & \textbf{Range} \\
    \hline
      \textit{client} & STOMP.js Client \\
      \textit{clientID} & \textit{UUID} \\
      \textit{pluginKey} & \textit{UUID} \\
      \textit{messageHistory} & [\textit{JsonNode}] \\
    \multicolumn{2}{c}{\centering \textbf{Methods}}\\
    \hline
    \textbf{Name} & \textbf{Return Value} \\
    \hline
    \textit{sendCreateMessage(JsonNode)} & \textit{bool}  \\
    \textit{sendUpdateMessage(JsonNode, UUID)} & \textit{bool} \\
    \textit{sendDeleteMessage(UUID)} & \textit{bool} \\
    \end{tabular}
\end{table}

Internally, the wrapper component defines a protocol, and spawns a thread for connecting to a websocket route on a predefined platform instance. Through subscribing to this websocket, all datagrams passed over instances of this plugin (identifiable by its \textit{pluginKey}) are deserialised and stored within the \textit{messageHistory}. This allows plugin designers to conditionally render the datagrams to their UI, based on the grammar they have defined. The wrapper also provides the three necessary methods for \textbf{CRUD} operations on the platform backend -- i.e., a \textit{create} method, taking as argument an arbitrary, serialisable datagram; an \textit{update} method, taking as arguments the ID of the datagram to change and the datagram to replace it with; and a \textit{delete} method, taking as argument the ID of the datagram to delete.

Accordingly, to comply with the \textit{React} paradigm, this set of methods is passed into the (dynamically injected) plugin component. This allows for the plugin developer, at design time, to use the pre-defined set of platform methods. This hides the inner functionality of the platform and simplifies the development process. We define this process as \textit{occlusion}, and describe it with greater detail in Section~\ref{sec:plugins}.

\subsection{Plugins} \label{sec:plugins}


Underpinning a bespoke, generative platform is its \textit{plugins}. In the context of \textit{PlatformOcean}, we leverage the \textit{React} framework to produce a dynamic, responsive web app; as such, the plugins must be compatible with this framework too. To achieve this, we model the plugins as a \textit{micro frontend} (MFE) -- i.e., decomposing the monolithic platform frontend into multiple smaller, self-contained frontends that can be dynamically injected and developed separately \cite{zateishchikov2023scaling}.

Given that the hosting platform is designed in React, it is only logical that the plugins should themselves be modelled as React \textit{micro-apps}: self-contained frontend units — like small, independently developed and deployed applications — that plug into a larger host app. By modelling them in such a way, each plugin can itself be a stand-alone React component and, using this React paradigm, accept arguments (or functions) into the component called \textit{props}.

Underpinning this plugin architecture is \textit{Module Federation's} dynamic remote component framework \cite{remoteComponents}. This framework allows for the dynamic inclusion of \textit{React} components from a foreign endpoint at runtime, when only the URL of the hosting platform is known. \textit{Module Federation} allows for pre-compiled React components to expose a \textit{remote entry} file, which can be injected into a host app (via a script tag) to register a component-providing module to the DOM's (document object model's) window object. From here, the component can be retrieved according to a pre-defined name and scope (declared during the bundling process).

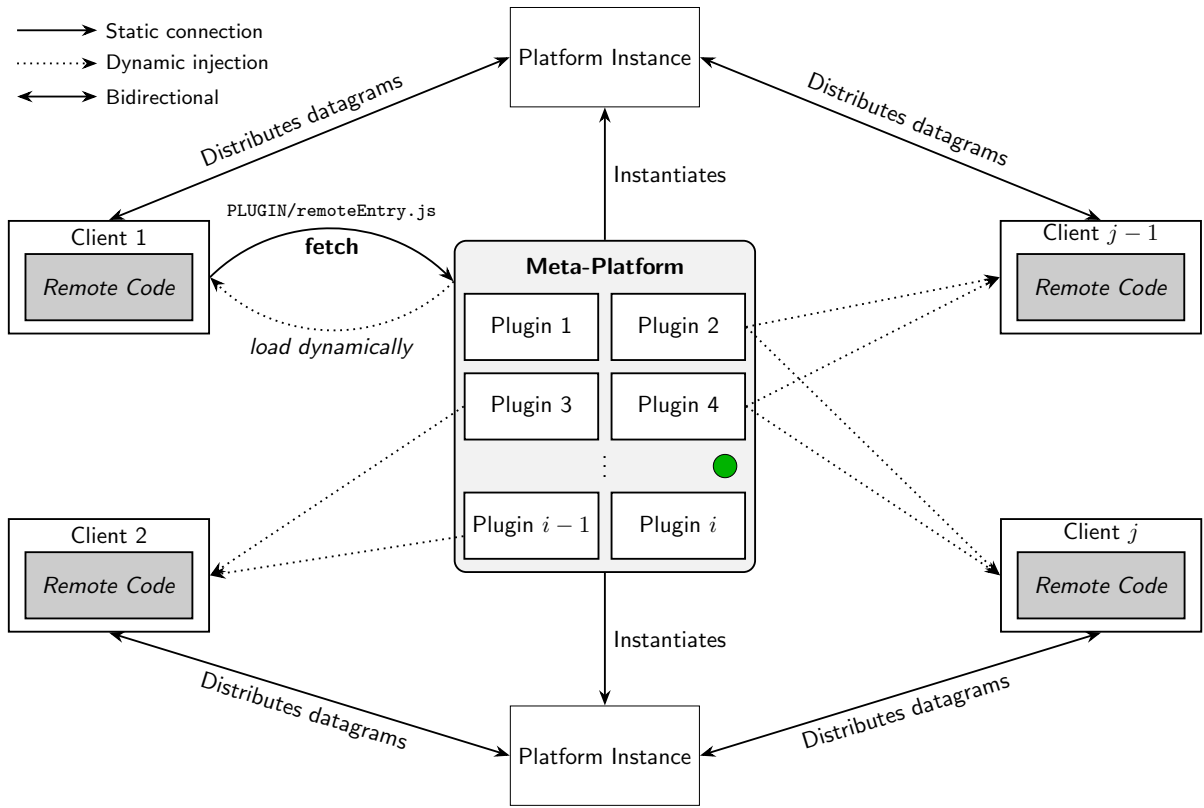
\begin{figure}[htb]
\centering
\resizebox{\linewidth}{!}{\input{tikz_plots/pluginArch.tikz}}
\caption{Full plugin architecture. Multiple clients connect to multiple servers, and dynamically inject multiple plugins}\label{fig:pluginArch}
\end{figure}


Achieving an effective, generic plugin architecture requires consideration of the performance of the system as a whole, according to a set of design criteria. For example, these plugins must be able to be dynamically included and removed without impacting the uptime of the platform. Furthermore, these plugins must be `immunised' against malicious intent, without compromising the genericity outlined in Section~\ref{sec:metaPlat}. Next, we examine these four criteria in more detail.

\subsubsection{Genericity}

It is impossible for us, the architecture designer, to encompass all possible use-cases of a developer's plugin. A key consideration for this system, then, is the capacity for \textit{genericity}, or the state of a plugin to be totally generic. A convenient way to achieve this is to enable the use of other libraries in the design process, such that the final plugin can leverage the entire library database, over a reduced set of methods supplied by the architecture.

Genericity also means that the developer should be able to make full use of the language's native support. For example, plugins developed in React should be dynamically responsive to changes in data through the \textit{useEffect} function, and allow for state maintenance through the native \textit{useState} function. This prevents developers from having to learn a new paradigm to align with the platform provider, and can instead rely on familiarity from the inherent features of the language.

This design criterion is addressed by using Webpack, wherein a dependency graph is built to span the local plugin files, any imported third-party libraries, and the (recursive) dependencies of the libraries themselves. These dependencies are all codified, and written into a local file bundle in the same format as `conventional' JavaScript code. With this, it is possible to create any conceivable plugin, as it can be implemented using a minified React \textit{micro-app}.

\subsubsection{Occlusion}

Reductively, a plugin should be considered simply as a way of accepting and displaying data. The hosting platform can be considered (equally simplistically) as a `data relay' where input data is taken from one system, sent to a server, and redistributed to a set of other systems. In this paradigm, a plugin simply becomes a tool for handling data input, and rendering data output.

Therefore, the onus is on the platform itself to provide all prerequisite distributed data functionality, such that these methods can be called from within the plugin. We refer to this paradigm as \textit{occlusion}, as all `difficult' distributed system components are `occluded' by a simple set of a methods, which hide the underlying platform functionality.

Since all data-distribution methods defined on the platform are essentially HTTP (Websocket) calls, a higher-order `wrapper' component \cite{scott2023system} can be defined. This defines the methods used for interfacing with the platform, and passes them into the plugin component (via its \textit{props}) for the plugin designer to call in accordance with the API, thereby hiding all inner-workings of the platform, and addressing the design criterion.


\subsubsection{Hot-Plugging}

To align with the high frequency usage of a platform, this plugin architecture must be designed for maximal platform uptime (and therefore minimal downtime). Inclusion/downloading of a plugin can therefore not allow platform reloads or restarts, or otherwise long periods of inactivity. A design approach that facilitates this is \textit{hot-plugging} \cite{wolfinger2006component}, wherein plugins can be included and removed at \textit{runtime}, without restarting the supporting platform.

This criterion is addressed by using the \textit{Module Federation Plugin} for \textit{Webpack}, which allows for this \textit{micro-app} to be dynamically included in a hosting platform at runtime.

\subsubsection{Sandboxing}

To support a wide array of customisation, the platform must support multiple different plugins without unwanted crosstalk or corruption of the hosting software. To achieve this, plugins must be \textit{sandboxed} \cite{wolfinger2006component}, with respect to both each other and the hosting platform. This entails that plugins should be included within a secure, bounding environment to prevent (possibly Byzantine) corruption of the host or other independent plugins. Plugins are then safely run within well-defined limitation to their actions.

As opposed to directly injecting the plugin into the main DOM where it may interact with other elements in the DOM and be able to execute arbitrary code in the document window,  inject the plugin is injected into its own isolated environment. By using \textit{iframes} another HTML page is loaded within the document, essentially putting another webpage within the parent page. This keeps all plugin instances isolated from one another, and the main window, satisfying this design criterion.

\section{Toolchains}\label{sec:toolchains}

In addition to the architecture itself, and its implementation through the \textit{meta-platform}, \textit{platform}, and \textit{plugins} described in Section~\ref{sec:FSA}, a set of toolchains are also provided to assist both the \textit{developer} and \textit{end user} with ease-of-use of the self-organisation platform.
Accordingly, we describe the plugin-development toolchain in Section~\ref{sec:devTools}, and the platform-interfacing toolchain, complete with a tool for performing the platform discovery in Section~\ref{sec:userTools}.

\subsection{Developer Tools} \label{sec:devTools}

In order to satisfy the plugins' design criteria outlined in Section~\ref{sec:plugins}, whilst still ensuring that it is still compatible with the system, a \textit{PlatformOcean} plugin must remain within the guardrails defined by the platform. For example, since all plugins must be dynamically injected into the host system, they must satisfy a common protocol to minimise variability. Specifically, as the plugin is injected into the document window, a standard scope is required. Furthermore, the \textit{remoteEntry} meta-file must also be generated consistently.

Given the variety of pitfalls that can arise with specifying this common set of protocols, we provide a toolchain for developers to handle the structuring, bundling, and publishing of plugins. This tool is called the \textit{pluginBundler}, and all of its constituent tools are detailed below.

\subsubsection{Scaffolding}

Taking inspiration from the \texttt{create-react-app} package provided by the official \textit{React} documentation for web-app development, we provide the necessary boilerplate code required for producing a plugin for \textit{PlatformOcean}.

For plugin development, not only do we require a carefully constructed configuration file (for integration with webpack), but a pre-ordained file structure to implement it. To avoid any possible difficulties with library downloads, file structure and configuration files, we pre-package the relevant scripts into a binary file, and allow it to be injected into a folder of the user's choosing. This completely skips any (potentially code-breaking) setup, allowing the user to enter directly into the development phase.

\subsubsection{Editing}

The \textit{pluginBundler} also provides a \textit{dev} environment, allowing a developer to mock up the functionality of a plugin, as if it were to be injected into a real platform. This generates a local development environment compatible with \textit{React's hot module replacement} (HMR) for dynamic UI updates, served by localhost, and viewable in the browser.

The editing tool renders two instances of the plugin specified by the user. These two instances mock up the distributed functionality of the platform using local methods in order to aid plugin development.

\subsubsection{Bundling}

Once a developer is satisfied with their plugin, they can bundle it into a minified format (using the \textit{prod} environment) for serialisation and distribution. This command generates a \textit{dist} bundle, and subsequently zips it, for uploading to the \textit{meta-platform}.

\subsubsection{Publishing}

Having produced this minified bundle, the developer can then publish their plugin to the plugin repository in the \textit{meta-platform}. This tool has the user enter the name of their plugin into the console, and then confirm it. Upon confirming the name, a connection is established with the \textit{meta-platform} and the plugin is published to the repository (via a standard POST request to the server's endpoint).

\subsection{User Tools} \label{sec:userTools}



To serve both as a means of identifying potential clients, and for providing an extra layer of security, we formalise a protocol that all browsers and servers must obey to perform a successful service discovery. Defining a specific protocol means that external actors would need to prior knowledge of the protocol in order to make use of the service, as the server would otherwise not respond. By keeping this protocol secret, we therefore protect the users from attacks. This protocol is defined in Algorithm~\ref{alg:ssdp}.

\begin{algorithm}[htb]
\caption{SSDP protocol for \textit{MulticastSubscriber}}\label{alg:ssdp}
\begin{algorithmic}
\State $\textit{multicastSubscriber} \gets \textbf{*spawn new thread*}$
\State $\textit{buffer} \gets byte[256]$
\State $\textit{socket} \gets \textit{MulticastSocket()}$
\State $\textit{group} \gets \textit{InetSocketAddress()}$
\State $\textit{response} \gets \textbf{*serialise server information*}$
\While{true} \Comment{$\textit{MulticastSubscriber.run()}$}

\State $\textit{packet} \gets \textit{DatagramPacket(buffer)}$
\State $\textit{socket.receive(packet)}$

\If{\textit{packet.matches(\textbf{codeword})}}
\State $\textit{senderIP} \gets packet.getAddress()$
\State $\textit{unicast(senderIP, response)}$
\EndIf
\EndWhile
\State $\textit{socket.close()}$
\end{algorithmic}
\end{algorithm}

This protocol creates a `codeword' that is only known by the client and server. The client serialises this codeword and multicasts it across the multicast group. If the server receives this codeword, it responds to the client by unicasting its network information back. (At present, a fixed codeword is used, in future this can be randomised and/or encrypted for increased security.)

\section{Case Studies}\label{sec:caseStudies}

\begin{figure*}[htb]
    \centering
        \subfloat[\textit{Squad management} plugin window -- players in the squad can be added, removed, and (privately) rated by dragging into order]{
            \includegraphics[width=.22\linewidth,height=.22\linewidth]{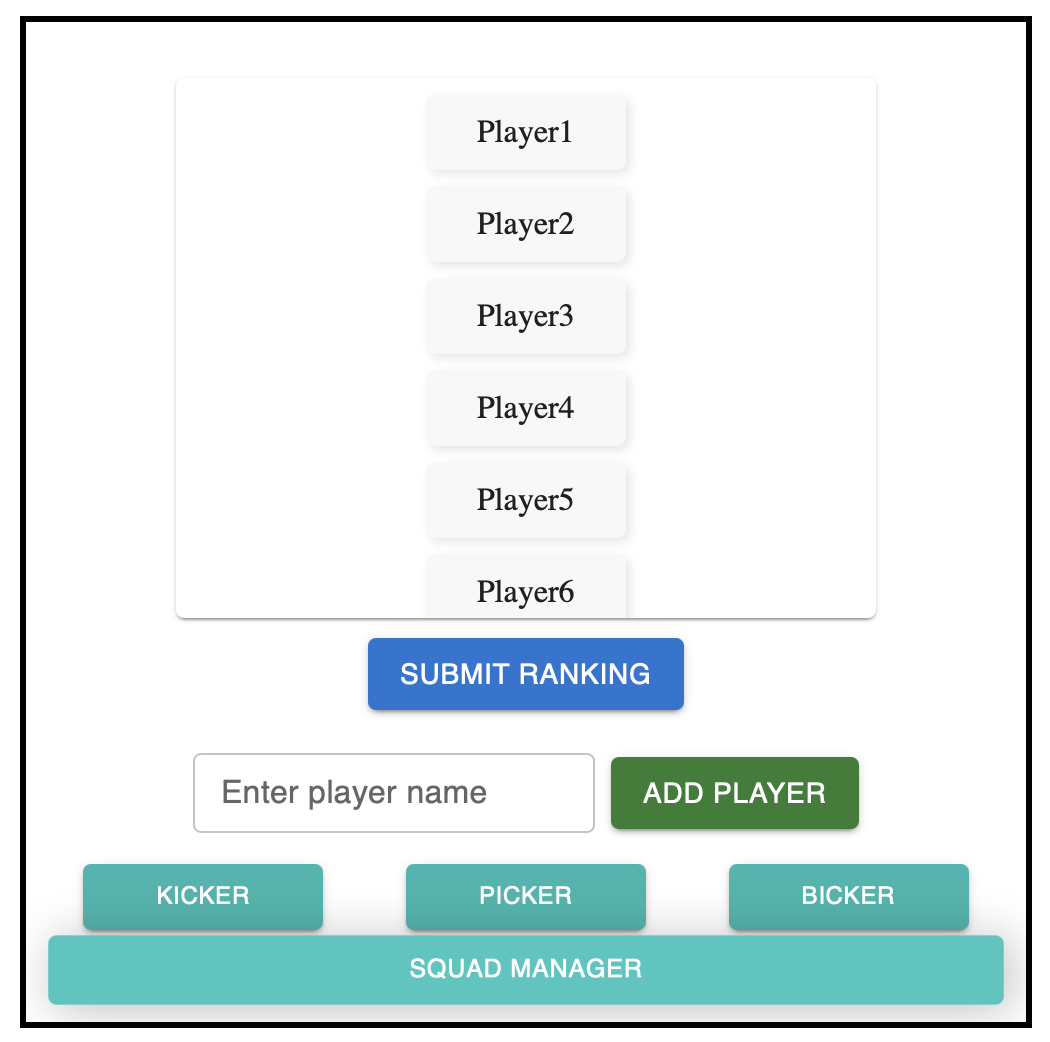}
            \label{fig:mananger}
        }
        \hspace{0.5em}
        \subfloat[\textit{Availability} plugin window -- the match is scheduled using a calendar, and availability is toggled by clicking the relevant name]{
            \includegraphics[width=.22\linewidth,height=.22\linewidth]{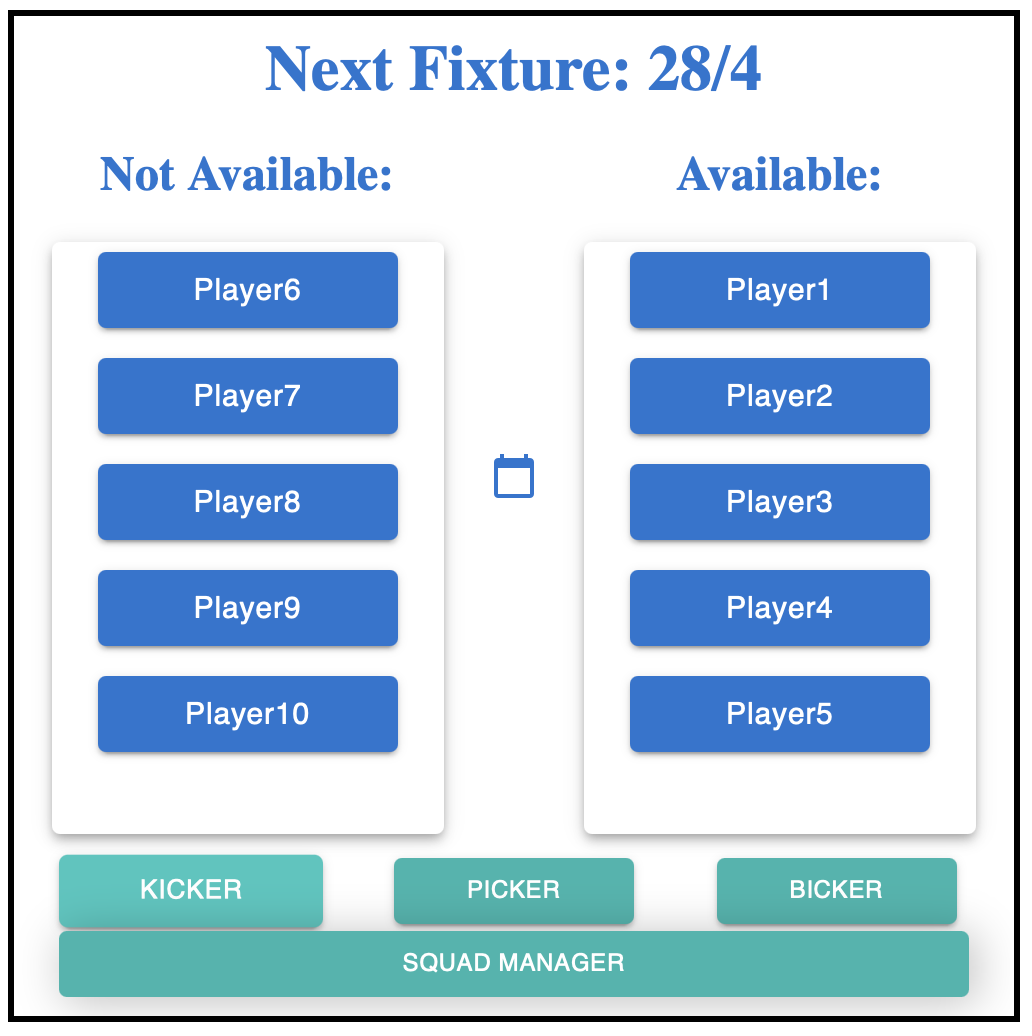}
            \label{fig:kicker}
        }
        \hspace{0.5em}
        \subfloat[\textit{Team recommendation} plugin window -- a list of potential matchups and their \textit{probability of fair game} (PFG) are generated for selection]{
            \includegraphics[width=.22\linewidth,height=.22\linewidth]{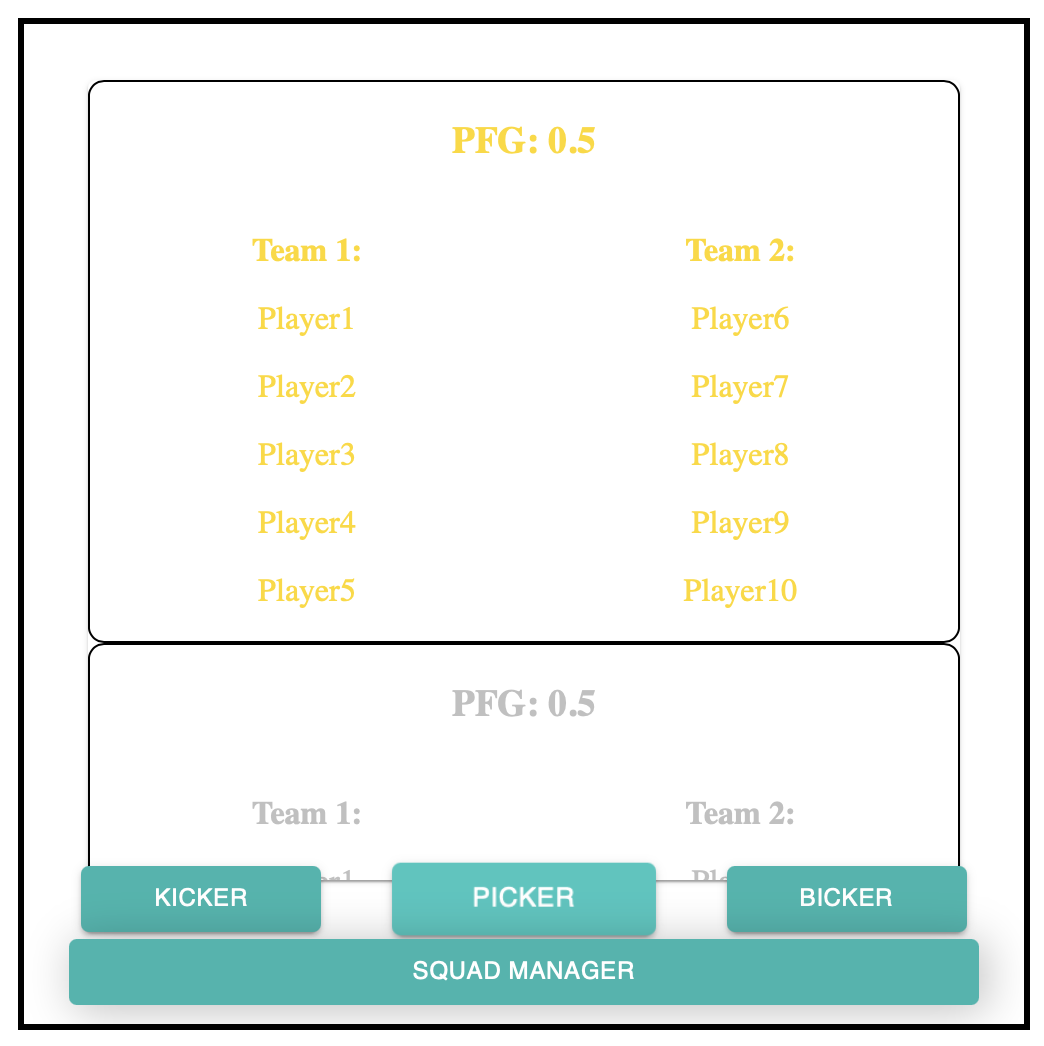}
            \label{fig:picker}
        }       
        \hspace{0.5em}
        \subfloat[\textit{Metrics} plugin window -- teammates rank the fairness of the game from 0.0 to 1.0, to help inform future matchups]{
            \includegraphics[width=.22\linewidth,height=.22\linewidth]{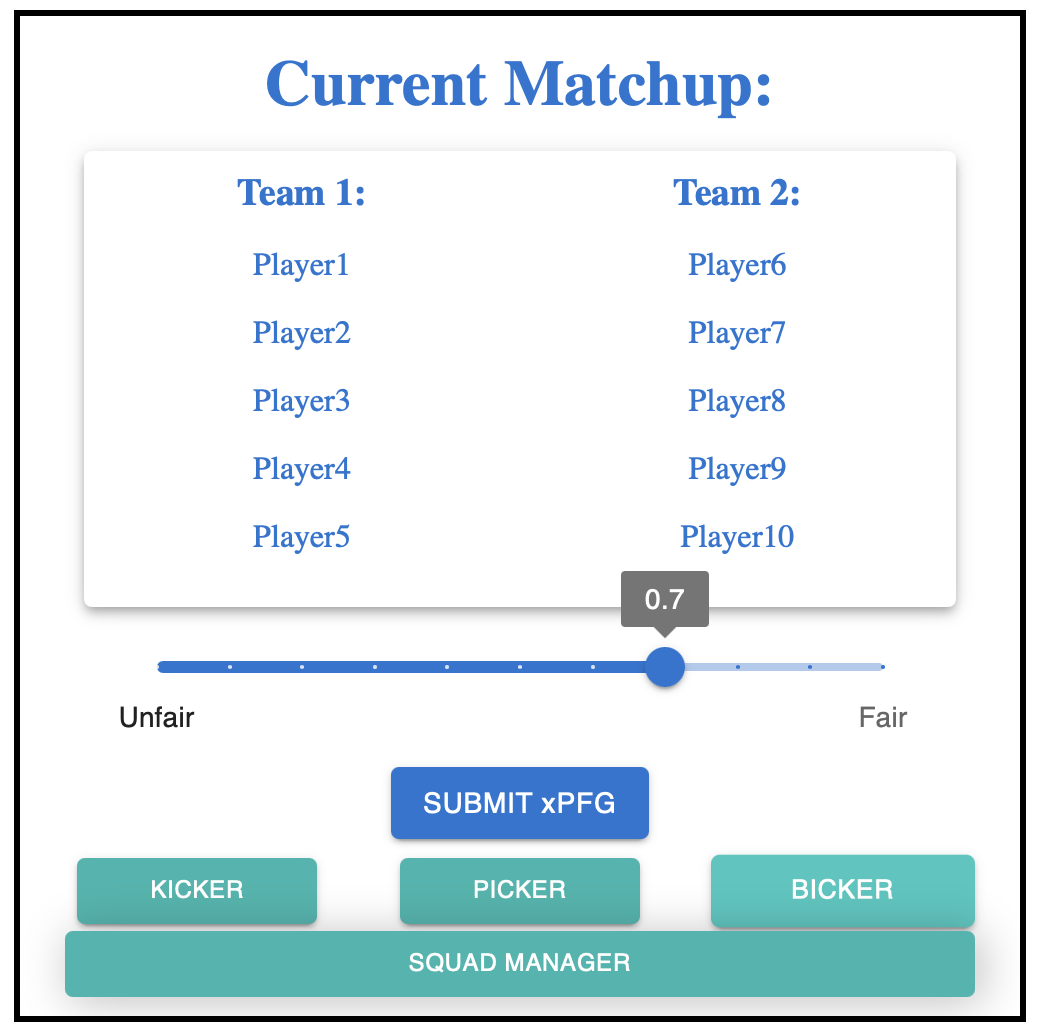}
            \label{fig:bicker}
        }

    \caption{Windows of plugin used to address the \textit{squad management}, \textit{team picker}, and \textit{management rater} functional requirements}
    \label{fig:squadPicker}
\end{figure*}

This section describes two application-specific platforms using the developer and user toolchains of Section~\ref{sec:toolchains} to engineer the full-stack architecture of Section~\ref{sec:FSA}.
Firstly, we describe the \textit{Sporting Association Coordinator}, a 
social coordination platform that assists a sports team to self-organise matches and coordinate various support activities (availability, 
team selection, travel, laundry, etc.). Secondly, we address the issue of
attention in education, and implement a collective study platform that 
enables a group of students to coordinate and incentivise their study habits.

As a proof-of-concept, we note how the two quite different applications have been
constructed from the same architecture using the same tools.
This demonstrates the scope for both bespoke personalisation and 
customisation, even for platforms of the same `type', and for seamless
re-use of plugins (both applications use a chat plugin, for example).
Moreover, both applications demonstrate how the architecture and toolchains
support community self-organisation at the application layer,
by design, in development and during operationalisation.




\subsection{Case Study I - Sporting Association Coordinator}\label{sec:hittenhope}

Hittenhope is a (genuine) football club that arranges an informal 5-a-side
football match once a week, at a regular time-slot at a local Soccer Centre 
(a private operator of dedicated football pitches for hire).
Teams are selected according to 
whoever is available that week; but a judicious selection is required to
ensure a `fair' and `balanced' game that is close and enjoyable for everyone.
Beyond notification of availability and team selection, there are various other activities associated with a game, such as travel and kit maintenance (i.e., laundry).


\subsubsection{Functional Requirements.}
An informal `player-centred' participatory design process identified the following functional requirements. Players should be able to\ldots

\begin{itemize}
    \item \textit{Squad management}: \ldots add and remove
    people from the squad (i.e., the pool of possible players);
    \item \textit{Availability}: \ldots notify that they
    are available for selection in a particular week;
    \item \textit{Team recommendation}: \ldots ask for 
    two `fair' teams to be automatically recommended,
    selected from those who have indicated their availability;
    \item \textit{Metrication}: \ldots rate each other privately,
    and rate specific games according to their personal perception of `fairness', 
    providing data for automated team recommendation;
    \item \textit{Communication}: \ldots engage in multi-logue `chats' as per any other social media instant messaging system;
    \item \textit{Participation}: \dots visualise fairness in `support'
    activities, such as kit maintenance (e.g., laundry); and
    \item \textit{Coordination}: \ldots minimise expense, emissions, and parking pressure, by forming car pools for available players.
\end{itemize}

Taking the meme ``there is an app for that'' seriously for a moment, each
one these separate functional requirements could be satisfies by a single app.
However, a far better approach would be to incorporate each disparate app
as a \textit{micro-app} that can be plugged in and integrated within an
over-arching platform. \textit{PlatformOcean's} plugin architecture specifically targeted at supporting such an approach: the design principles outlined in Section~\ref{sec:plugins} enable each required plugin to be dynamically added, allowing the platform development to be validated against the functional
requirements.

The plugin architecture also supports the platform developer for UX (user experience). Each of the functional requirements could produces a completely different UI and plugin grammar. The \textit{genericity} outlined in Section~\ref{sec:plugins}, and the plugin development tool described in Section~\ref{sec:devTools} help with this design process for ensuring
consistency of look/feel and operation across plugins.


\subsubsection{Implementation}

Given that the \textit{squad management}, \textit{availability}, \textit{team recommendation}, and \textit{metrics} requirements are inextricably linked, we implement all four within the single, \textit{tabbed} plugin shown in Figure~\ref{fig:squadPicker}.

The \textit{genericity} of the plugin architecture and development toolchain are both leveraged to produce this plugin. Most interface elements use the \textit{React} (MUI) framework, combined with a bespoke grammar for state management and data representation. This is achieved using the external library support from the \textit{Webpack bundler}, and arbitrary \textit{JsonNode} representation supported by the platform, respectively.

A plugin designer must consider the plugin grammar both from the perspective of sending and receiving datagrams. This grammar must be defined such that each component of this plugin is uniquely characterised, and the state is synchronised across all platform instances.

The \textit{squad management} component of the plugin has two functions - adding/removing players from the squad, and submitting a rating of these players. Accordingly, this plugin should produce two messages: firstly, a means of identifying a player. For reusability (to minimise repeated data in the platform), we also attach the player's availability to this datagram, so that it is useable in the \textit{teach recommendation} component:


This plugin component then dynamically re-renders in response to the message. This particular implementation defines a custom \textit{React hook} to extract the relevant messages from the message history, and maps them from a list of datagrams to a list element.

Secondly, this component must produce messages that represent the rating of a squad. Again, we leverage the arbitrary \textit{JSON} format to uniquely characterise the message type, and use a list of objects to codify the ranking:


\noindent which is also compatible with the custom \textit{React hook} described above, allowing the \textit{team recommendation} plugin to prefetch the relevant messages, preprocess them for input to the matchup generation algorithm, and render the output.


\begin{figure*}[htb]
    \centering
        \subfloat[\textit{Communication} plugin -- messages are conditionally rendered based on the active user. Arrows below the plugin toggle chat windows]{
            \includegraphics[width=.22\linewidth,height=.22\linewidth]{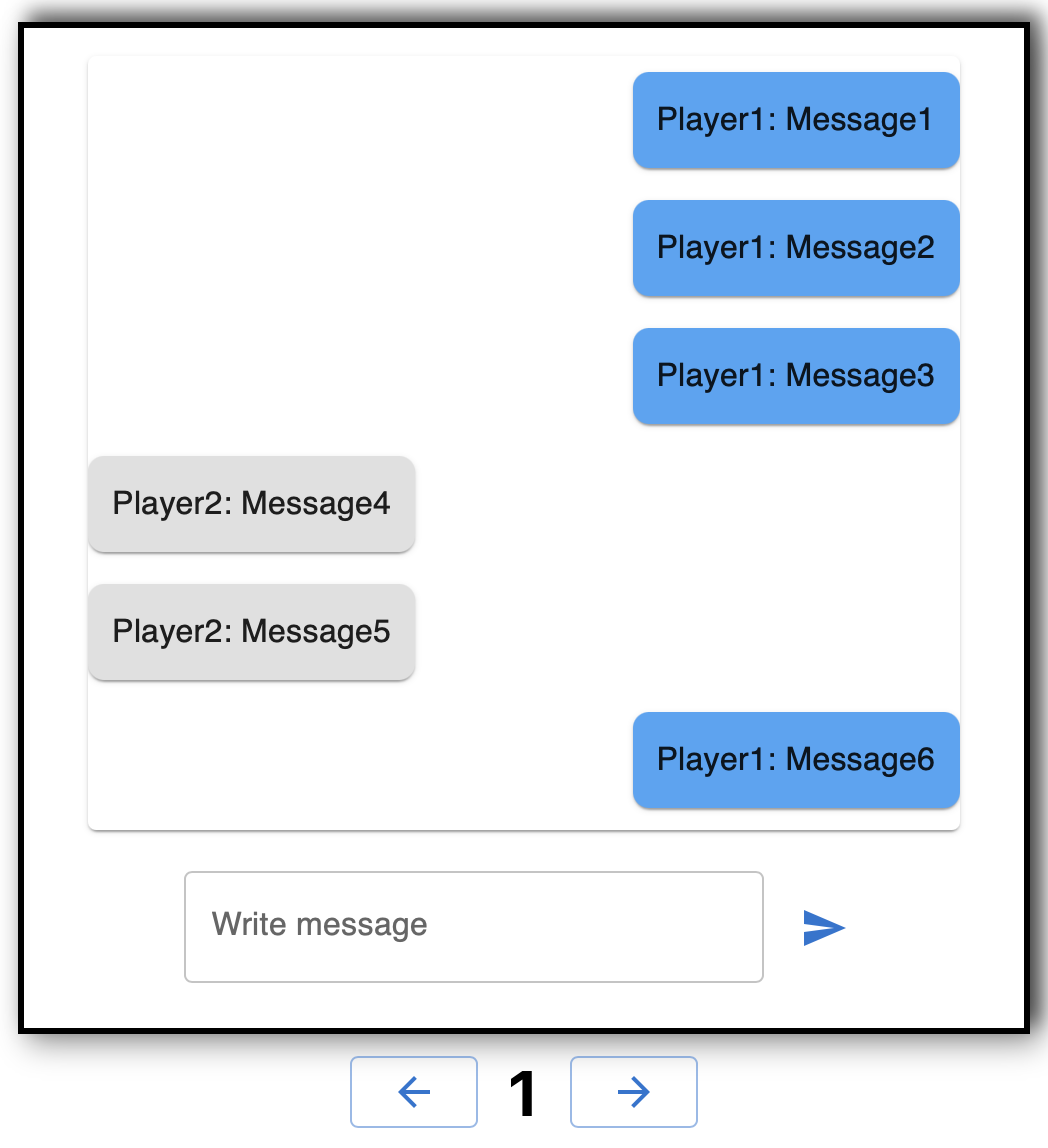}
            \label{fig:notSap}
        }
        \hspace{0.5em}
        \subfloat[\textit{Equal participation} plugin -- a colour-coded array of names maps the player's laundry count to `hotness' in a heatmap]{
            \includegraphics[width=.22\linewidth,height=.22\linewidth]{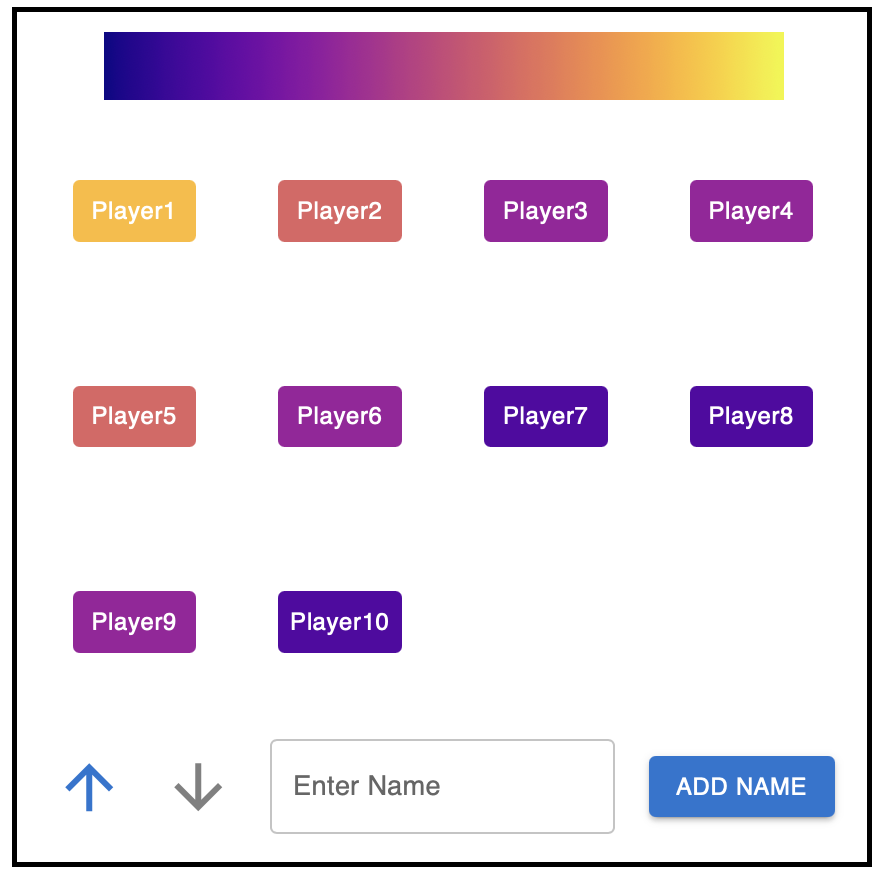}
            \label{fig:laundryRota}
        }
        \hspace{0.5em}
        \subfloat[\textit{Coordination (1)} plugin -- players upload a name and postcode to form a list of reorderable waypoints. First name identifies the designated driver]{
            \includegraphics[width=.22\linewidth,height=.22\linewidth]{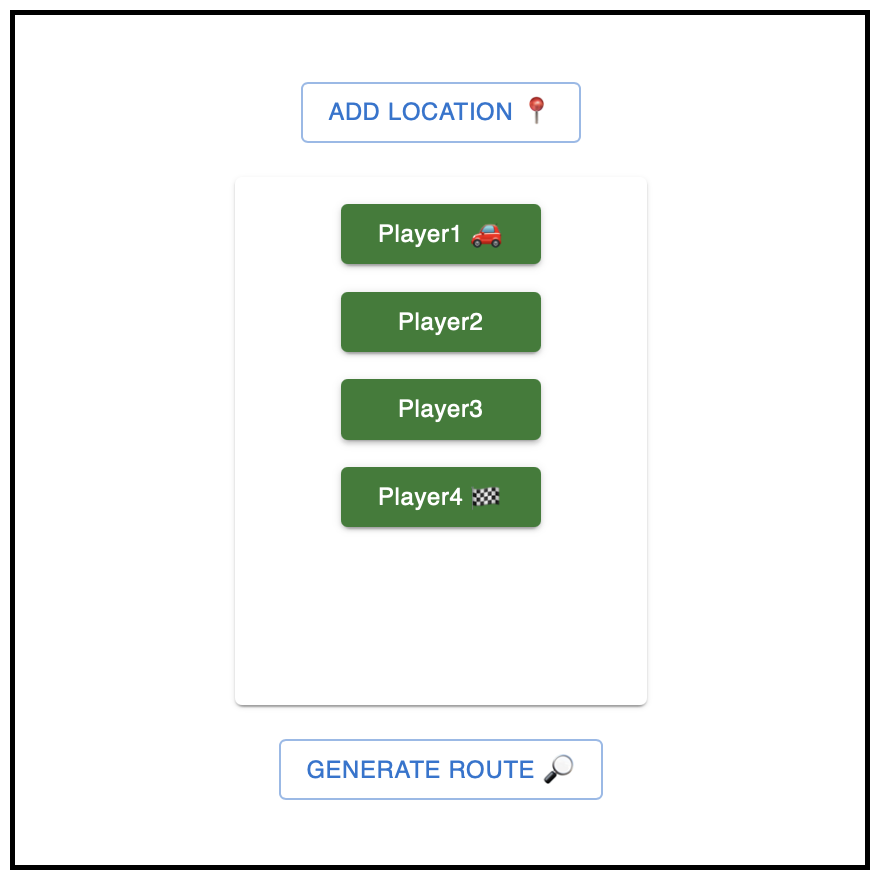}
            \label{fig:carpooler}
        }
        \hspace{0.5em}
        \subfloat[\textit{Coordination (2)} plugin -- output of the \textit{coordination} plugin route generation. An interactive map from a free API is embedded in an \textit{iframe}]{
            \includegraphics[width=.22\linewidth,height=.22\linewidth]{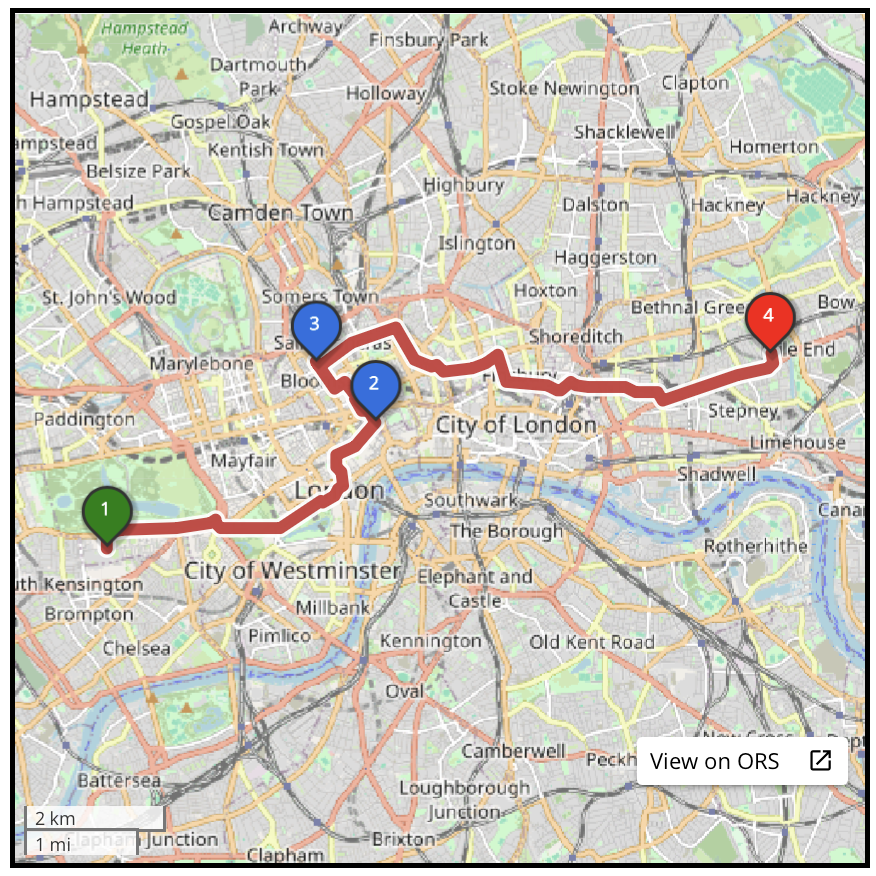}
            \label{fig:mapper}
        }

    \caption{Plugins used to address the \textit{chat app}, \textit{laundry rota}, and \textit{carpooler} functional requirements}
    \label{fig:podsQuicker}
\end{figure*}

Similarly, we provide the implementation of the \textit{chat app} in Figure~\ref{fig:notSap}. This plugin leverages the functionality of the wrapper component \cite{scott2023system} to dynamically inject the name of the registered user into the sent datagram. This history of datagrams is then conditionally rendered to the UI, where a grey, left-aligned styling is applied to messages that originate from a different user instance (identifiable by the \textit{sender} parameter in Table~\ref{tab:messageSchema}), and a blue, right-aligned styling otherwise.


Furthermore, we address the `fair' \textit{participation} functional requirement with the custom plugin shown in Figure~\ref{fig:laundryRota}. To codify this requirement, we envision a list of datagrams comprising a mapping of player name to laundry count. Here, the players' names can be entered and removed to display a square grid, with each name superimposed on a button with colour (`heat') proportional to the frequency of laundering.


Finally, to address the \textit{coordination} functional requirement, we provide a custom car-pooling plugin in Figure~\ref{fig:carpooler}. This plugin allows players to enter their name and address to produce a waypoint. This list of waypoints is then rendered to the UI as a \textit{draggable}, such that users can identify the start and end locations of the route. Pressing \textit{generate route} submits a request to the \textit{Open Route Service} API, to produce a route.


\subsection{Case Study II - Group Study Supporter}\label{sec:mindMyMind}

In the context of the data economy \cite{zuboff2023age},
the idea of privacy as the right not to be observed has led
to the development of PET (Privacy Enhancing Technology).
More recently, in the
context of the so-called attention economy, the
idea of attention as the right not to be interrupted has led to
analogous proposals for Attention Enhancing Technology \cite{aet}.
This is especially important for contemporary students, who have to focus
on coursework and examinations when an induced dependence on mobile devices and
social media, distracting their focus and siphoning their cognitive energy,
is a deliberate factor in product and service design.
Therefore, we propose to develop a platform
for personalised, self-organised, mutually-supportive study that
promotes concentration and preserves attention.

\subsubsection{Functional Requirements}

The primary motivation for the Group Study platform is
support for collective study, where small groups of students can coordinate
their time according to their preferred study method.
An informal `student-centred' participatory design process identified the following functional requirements. Students should be able to\ldots
\begin{itemize}
    \item \textit{Selection}: \ldots adjust a dynamically-adaptive task management systems according
    to their own diverse preferences and study habits;
    \item \textit{Feedback}: \ldots exploit `internal' mechanisms for incentivising themselves and
each other, whereby study is its own immediate reward, but provides
longer-term reward in improved performance and grades; and
    \item \textit{Reward}: \ldots enjoy the benefits of 
    an 
`external' reward system that provides feedback and positive reinforcement. 
\end{itemize}
From the platform developers perspective,
techniques to satisfy these demands
include the use of the inner voice
\cite{innervoice} for its role in productivity and cognitive functions such as memory, decision-making, and emotional regulation; and a gamified system
leveraging 
Bartle’s Taxonomy of Player Types \cite{bartle}.
This taxonomy classifies gamers into four archetypes: \textit{achievers} (motivated
by rewards), \textit{explorers} (motivated by discovery or statistics), \textit{socialisers} (motivated by connection),
and \textit{killers} (motivated by competition).

However, this presents a different challenge to the previous case study, 
as the mapping between functional requirements and plugins is no longer
one-one. This demands that the implementation requires multiple plugins
for each requirement, in particular to support the full range of gamer types.

\subsubsection{Implementation}

\begin{figure*}[htb]
    \centering
        \subfloat[\textit{Task manager} plugin -- students create a list of tasks that can be marked as complete, deleted, and entered into \textit{focus mode}]{
            \includegraphics[width=.215\linewidth]{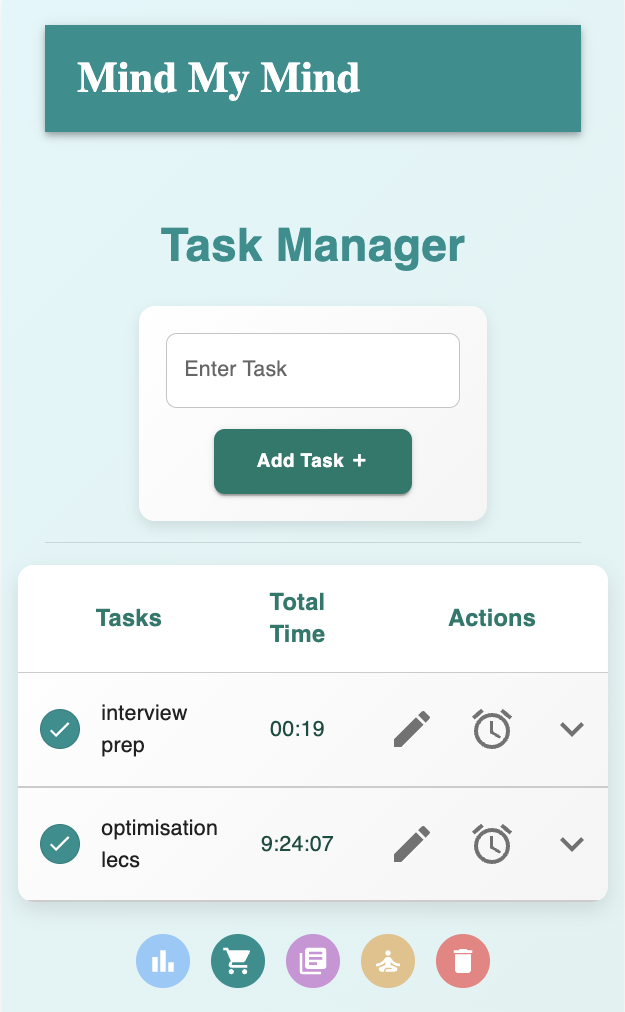}
            \label{fig:taskManager}
        }
        \hspace{0.5em}
        \subfloat[\textit{Study preferences} plugin -- students fill out a form about various study techniques to receive a recommended study method]{
            \includegraphics[width=.218\linewidth]{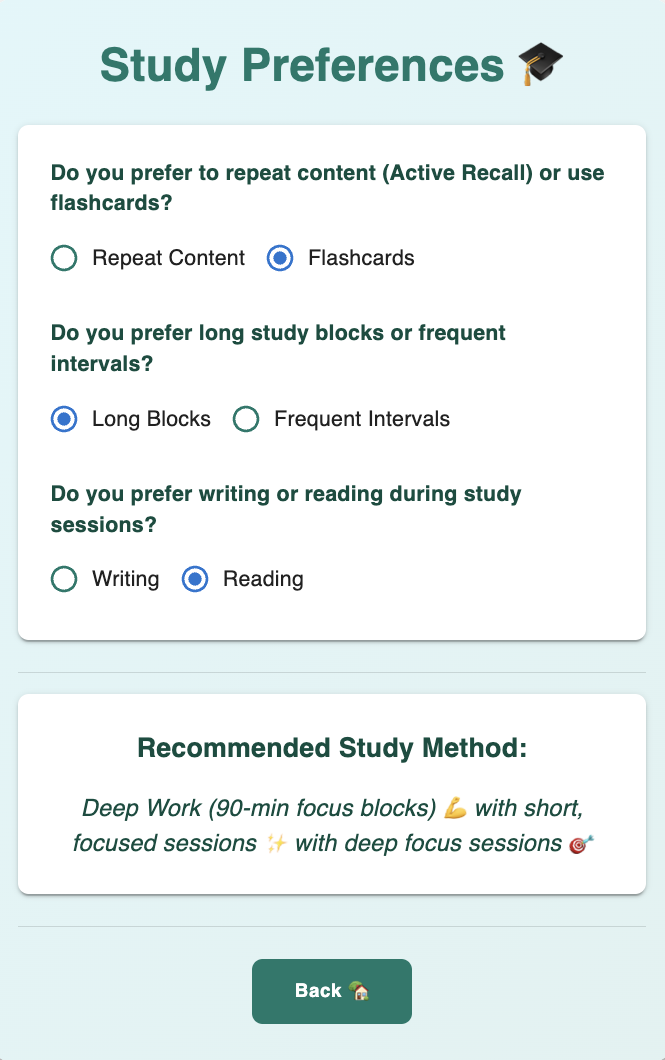}
            \label{fig:studyPrefs}
        }
        \hspace{0.5em}
        \subfloat[\textit{Statistics} plugin -- the results from the \textit{task manager} plugin are visualised to show study times per task and study trends]{
            \includegraphics[width=.225\linewidth]{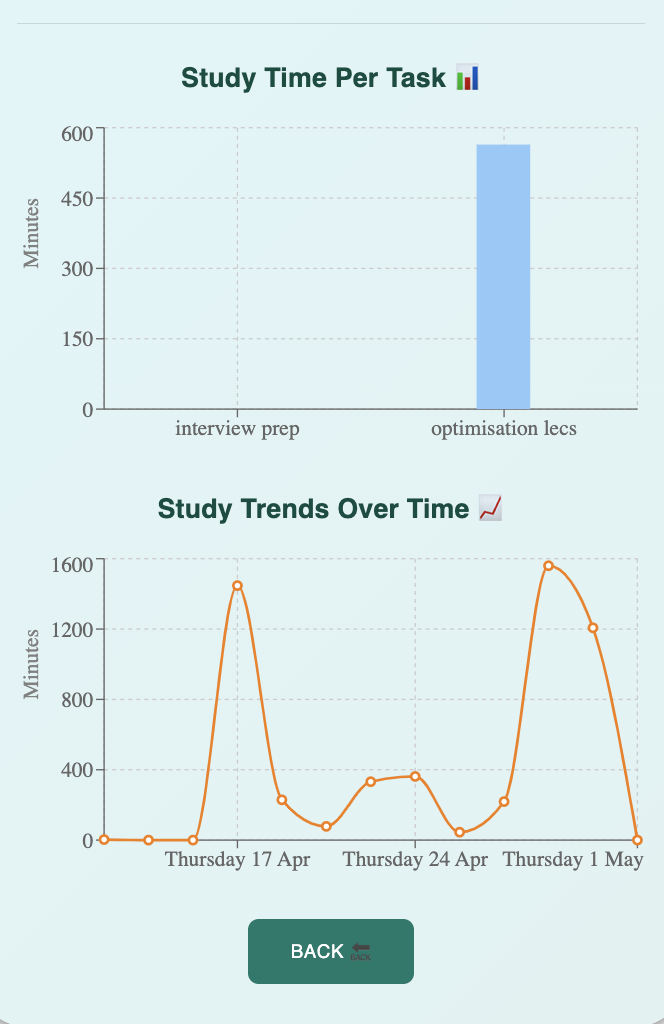}
            \label{fig:studyStats}
        }
        \hspace{0.5em}
        \subfloat[\textit{Motivational boost} plugin -- students are given randomised motivational quotes and allowed to self-reflect to soothe the \textit{inner voice}]{
            \includegraphics[width=.22\linewidth]{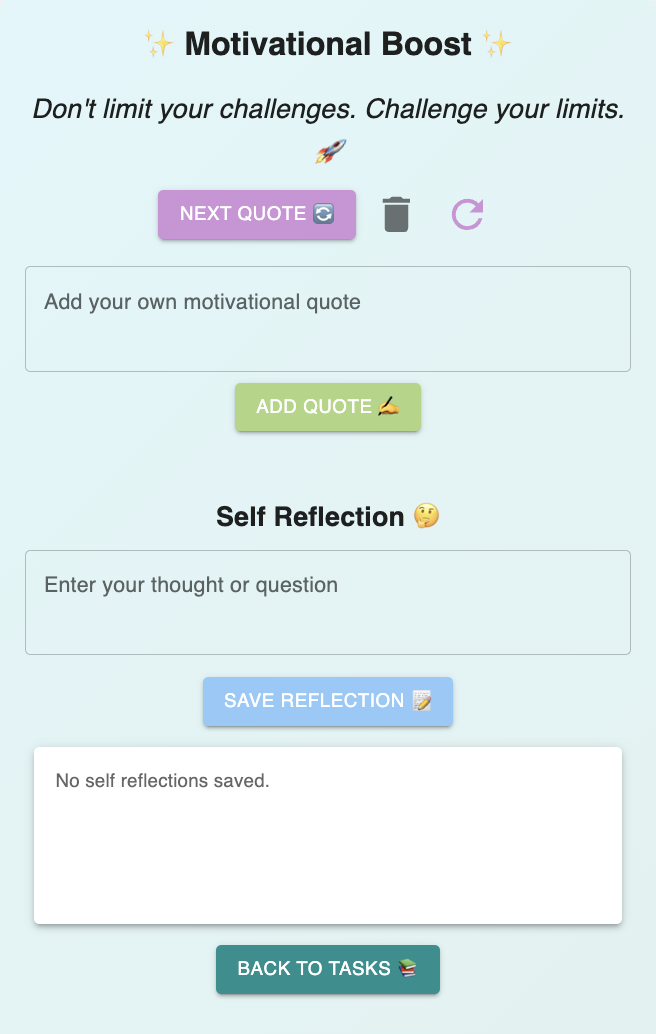}
            \label{fig:motivBoost}
        }
    \caption{\textit{Task manager}, \textit{study preferences}, \textit{statistics}, and \textit{motivational boost} plugins for the \textit{Group Study Coordinator} platform}
    \label{fig:mindMyMind}
\end{figure*}

As a baseline requirement, the \textit{Group Study Supporter} platform acts as as task manager for coordinating the students' time. Figure~\ref{fig:taskManager} shows the UI of the \textit{task manager} plugin. Here, students enter the name of the task into the form, whereby it is added it to a list of outstanding tasks. This list provides each task with its personal timer, showing how long the task has been underway, and a set of further actions, for controlling the state of a task -- i.e., editing the name of, or deleting the task. Critically, each task comes with a \textit{focus mode} (represented in Figure~\ref{fig:taskManager} by the clock icon), wherein the student is taken to a minimalistic, timer-based UI to record how long they have engaged in uninterrupted study. This promotes focus by removing external distractions.

In conjunction with the \textit{task manager}, the \textit{Group Study Coordinator} provides a \textit{study preferences} plugin to help customise the interface of the \textit{task manager}, based on a set of preferences and study habits. Figure~\ref{fig:studyPrefs} illustrates the form used to record user feedback in tailoring the \textit{task manager} interface. This satisfies the \textit{selection} functional requirement.

Furthermore, a mechanism for recording a student's progress is provided by the \textit{statistics} plugin, shown in Figure~\ref{fig:studyStats}. Here, a histogram records the study time per task, and a line graph records the average time spend studying each day to identify trends in students' studying habits. This mechanism addresses the \textit{feedback} functional requirement, and particularly supports the \textit{explorer} archetype, who are motivated by the observation and discovery of trends and statistics.

Finally, to address the \textit{reward} functional requirement, wherein students are provided with positive reinforcement for their studying,  Figure~\ref{fig:motivBoost} shows the \textit{motivational boost} plugin. Figure~\ref{fig:motivBoost}. The role of this plugin is twofold: firstly, inspirational quotes are randomly generated (from a database) to boost the students' morale. Secondly, the \textit{motivational boost} plugin calms the inner voice by acting as a journal, allowing students to record thoughts for self-reflection.

Beyond the `core' components shown in Figure~\ref{sec:mindMyMind}, we also provide a repository of further plugins to help customise a bespoke study platform. In particular, there are plugins for quizzes to get feedback to appeal to \textit{explorer} types. There is also a chat-app plugin and kudos system to appeal to \textit{socialiser} types. Here we can re-use the chat-app from Section~\ref{sec:hittenhope}, demonstrating how different platforms can exploit the same plugins albeit for different purposes. To appeal to the \textit{achiever} type, a plugin can be added to let students earn `coins' (a conceptual, gamified currency) to spend in a `shop', giving options for customisation, say. Finally, to appeal to the \textit{killers}, we provide a leaderboard plugin which records the data from the \textit{statistics} plugin and compares it against other students' performances, to encourage competition.

A platform can be defined as a set of components. This set comprises low-variety, core components and high-variety peripheral components. The low-variety core provides the base functionality and is relatively stable, while the high-variety peripherals offer customization and variation \cite{baldwin2009architecture}. In this paradigm, the \textit{task manager} and \textit{motivational boost} plugins are low-variety, core components, which are instantiated on the `vanilla' \textit{Group Study Supporter} platform. The \textit{statistics}, \textit{chat-app}, \textit{quizzes}, and \textit{leaderboards} plugins, then, serve as the high-variety \textit{complementary} components, allowing for bespoke customisation of the pre-instantiated platform.

\section{Related and Further Work}\label{sec:furtherWork}

Underpinning \textit{PlatformOcean} is its design as a \textit{plugin architecture}; particularly one that supports the development of generic plugins in \textit{JavaScript}, using the \textit{React} paradigm. One implementation of this is the \textit{pluggy-nx} library \cite{reactPluginArch} where, similarly to \textit{PlatformOcean}, \textit{module federation's remote modules} are used to dynamically inject code into a host app, from a foreign source. This implementation similarly provides a simple toolchain for bundling and publishing these plugins to a repository. Where \textit{PlatformOcean} differs, however, is with respect to 1. the complexity of the toolchains, and 2. the `higher-level' requirements of the plugin architecture. Producing plugins in this paradigm requires well-defined guardrails -- both through the generation of the remote entry \textit{metafile} and the structure of the plugin itself. This is solved by providing developers with the developer toolchain in Section~\ref{sec:devTools}, to \textit{scaffold} the required plugin structure, and subsequently \textit{bundle} the plugin (along with its remote entry file) into a distributable. 

Furthermore, \textit{PlatformOcean} takes the plugin architecture a step further to produce a \textit{distributed} social-coordination platform. This redefines the plugins as vehicle for sending, receiving and distributing arbitrary datagrams, using the protocol defined on the platform. As such, we introduce the novel \textit{wrapper} component from Table~\ref{tab:pluginAtts} to dynamically inject this functionality into the plugins, and achieve the design requirement of \textit{occlusion}, outlined in Section~\ref{sec:plugins}. The platform is also dynamically responsive to the addition/removal of plugins, achieving \textit{hot-plugging}.

Another alternative to global digital platforms for social
coordination is the grassroots architecture \cite{shapiro2024grassroots}.
The three core concepts of grassroots
architecture are: digital sovereignty, which seeks
to construct digital analogues of physical notions of
control over space, currency, and data;
grassroots systems, which are independent,
but interoperable, distributed serverless systems, relying only on compute power of edge devices (such
as smartphones) and not on global resources; and
blocklace, an extension of the blockchain data
structure, which allows for cooperative convergent
(re)construction without global coordination and
consensus. Grassroots and \textit{PlatformOcean} are clearly
aligned on the first two conceptual points, although the integration
of blockchain technology for axial currencies in \textit{PlatformOcean}
is an issue for further work.

Other ongoing work includes evaluation of both case studies through user trials.
This has two dimensions: evaluation from a pure usability perspective, and
evaluation from a self-organisation perspective, especially if other user groups observe the platform in use and instantiate a platform for themselves. 
Although the \textit{Sporting Association Coordinator} includes some machine learning
from data provided by the team recommendation ratings, there is an opportunity
to develop an LLM-based plugin for the \textit{Group Study Supporter}, and this is 
currently being investigated. In addition, a third application is being developed in the
domain of a platform for school meal menu planning. Since different districts
have varying requirements for nutritional recommendations and food sources, platform re-use and customisation is essential. 

\section{Summary and Conclusions}\label{sec:sumConc}

In summary, this paper has described a full-stack architecture for platform
development, discussed developer- and user-oriented toolchains to support design and implementation, and demonstrated two proof-of-concept case studies, one in
social coordination and another in collective study.
The specific contribution of this work has been to show that  
self-organisation at the application layer can be achieved by the specific supporting functionality of a full-stack architecture with complimentary
developer and user toolchains for high and low variety plugins.
The significance of this work is to have demonstrated the feasibility of the
democratisation of platform technology for decentralised, self-organised
social coordination, thereby offering a viable alternative to existing
monopolistic practices \cite{pittnowaketal}.

In conclusion, the process of platformisation has resulted in
the unrestricted spread of digital platforms into every domain of human social activity. In the `analogue' world, these social interactions generally did not require supervision by a third-party middleman or interlocutor, but as they became increasingly computer-mediated, many forms of social coordination became correspondingly dependent on the intercession of various platforms.
Moreover, preferential attachment and centralisation at the application layer of the internet have meant that the initial proliferation of platform offerings has been mostly consolidated into a platform monopoly for each particular type of social coordination.  There follows significant potential for commercialisation, commodification and monetisation, at deep social cost.
\textit{PlatformOcean} offers a viable alternative, showing that it is technologically and practically possible for such platforms to be re-used, recycled, reformed -- and re-imagined: a continuous cycle of user-empowered self-improvement.

Just as a Linux distribution includes the kernel, supporting software and driver libraries -- most of which are provided by third parties -- a meta-platform instantiation includes the base platform, supporting software and customised plugins -- most of which would ideally be provided by
third parties. It can even be envisaged that there are platforms instantiated
specifically for the purpose of collectively producing plugins. In this way,
we could achieve for platformisation at the application layer
what Linux achieved for operating systems lower down the computer architecture
stack:
a customisable open-source platform underpinned by third-party open-source
plugins for unrestricted self-organised social coordination.


\bibliographystyle{ieeetr}
\bibliography{main.bib}

\end{document}

%% file: tikz_plots/p1.tex
\begin{tikzpicture}
\draw (0,0) -- (2,0) -- (2,2) -- (0,2) -- cycle;
\pgfmathsetmacro{\l}{0.3};
\pgfmathsetmacro{\x}{0.25};
\foreach \i in {0,2,4}{
    \foreach \j in {0,2,4}{
        \pgfmathsetmacro{\xone}{\x + (\l * \i)};
        \pgfmathsetmacro{\yone}{\x + (\l * \j)};
        \pgfmathsetmacro{\xtwo}{\xone + \l};
        \pgfmathsetmacro{\ytwo}{\yone + \l}
        \draw (\xone, \yone) -- (\xtwo, \yone) -- (\xtwo, \ytwo) -- (\xone, \ytwo) -- cycle;
    }
}
\node[anchor=center] at (1,1) {\includegraphics[width=0.35cm, height=0.35cm]{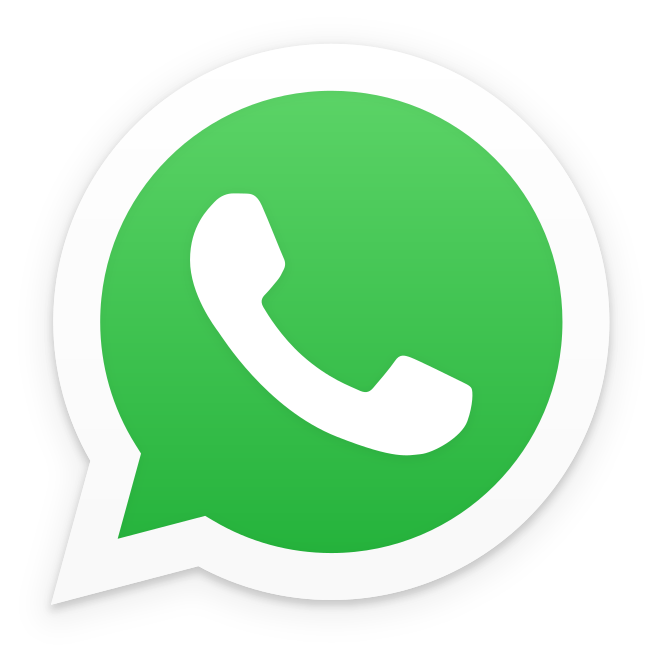}};
\draw[line width=0.75, color=red] (1, 1) circle (\l);  

\draw [dotted, line width = 1] (1, 1.3) -- (3, 2);
\draw [dotted, line width = 1] (1, 0.7) -- (3, 0);

\pgfmathsetmacro{\n}{4}
\pgfmathsetmacro{\dh}{2 / \n}
\draw (3,0) -- (5,0) -- (5,2) -- (3,2) -- cycle;
\foreach \i in {1,...,\n}{
\pgfmathsetmacro{\h}{\i * \dh}
\draw (3,\h) -- (5,\h);
\pgfmathsetmacro{\hcent}{\h - (\dh / 2)}
\draw[fill=gray] (3.25, \hcent) circle (0.1);  
\pgfmathsetmacro{\idx}{int(5 - \i)}
\node at (4.18, \hcent) {\fontsize{5.5}{2}\selectfont Group Chat \idx};
}

\draw [dotted, line width = 1] (5, 1.25) -- (6, 2);
\draw [dotted, line width = 1] (5, 1.25) -- (6, 0);

\draw (6,0) -- (8,0) -- (8,2) -- (6,2) -- cycle;
\node at (7, 1.8) {\fontsize{6}{2}\selectfont Group Chat $k$};
\draw (6, 1.65) -- (8, 1.65);
\pgfmathsetmacro{\n}{6}
\pgfmathsetmacro{\dh}{1.6 / \n}
\foreach \i in {1,...,\n}{
\pgfmathsetmacro{\h}{2 - (\i * \dh + \dh / 2) - 0.12}
\pgfmathsetmacro{\pos}{6.6}

\ifnum\i=4
  \pgfmathsetmacro{\pos}{7.4}
\else
  \ifnum\i=5
    \pgfmathsetmacro{\pos}{7.4}
  \fi
\fi

\node at (\pos, \h) {\fontsize{5}{2}\selectfont Message \i};
}

\draw (3,3) -- (5,3) -- (5,5) -- (3,5) -- cycle;

\draw[<->, line width = 1] (3,4) -- (1,2);
\draw[<->, line width = 1] (4,3) -- (4,2);
\draw[<->, line width = 1] (5,4) -- (7,2);
\node at (4, 4.6) {\fontsize{5}{2}\selectfont Centralised Server};
\node[anchor=center] at (4,3.75) {\includegraphics[width=1cm, height=1cm]{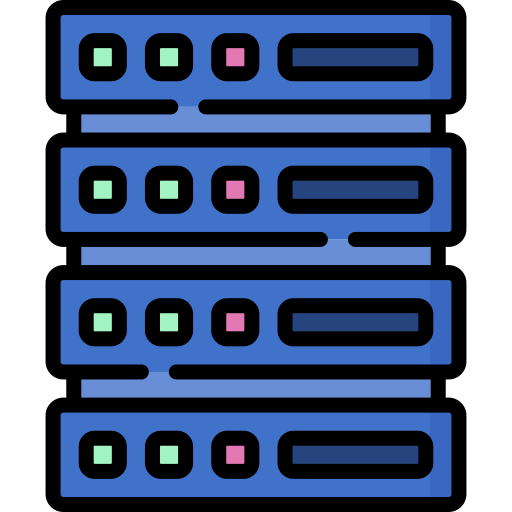}};

\end{tikzpicture}

%% file: tikz_plots/p2.tex
\begin{tikzpicture}

\draw [dotted, line width = 1] (-1, -1) -- (9,-1);

\end{tikzpicture}

%% file: tikz_plots/p3.tex
\begin{tikzpicture}

\draw (0,-4) -- (2,-4) -- (2,-2) -- (0,-2) -- cycle;
\pgfmathsetmacro{\l}{0.3};
\pgfmathsetmacro{\x}{0.25};
\foreach \i in {0,2,4}{
    \foreach \j in {0,2,4}{
        \pgfmathsetmacro{\xone}{\x + (\l * \i)};
        \pgfmathsetmacro{\yone}{\x + (\l * \j)};
        \pgfmathsetmacro{\xtwo}{\xone + \l};
        \pgfmathsetmacro{\ytwo}{\yone + \l}
        \draw [] (\xone, \yone - 4) -- (\xtwo, \yone - 4) -- (\xtwo, \ytwo - 4) -- (\xone, \ytwo - 4) -- cycle;
    }
}
\node[anchor=center] at (1,-3) {\includegraphics[width=0.35cm, height=0.35cm]{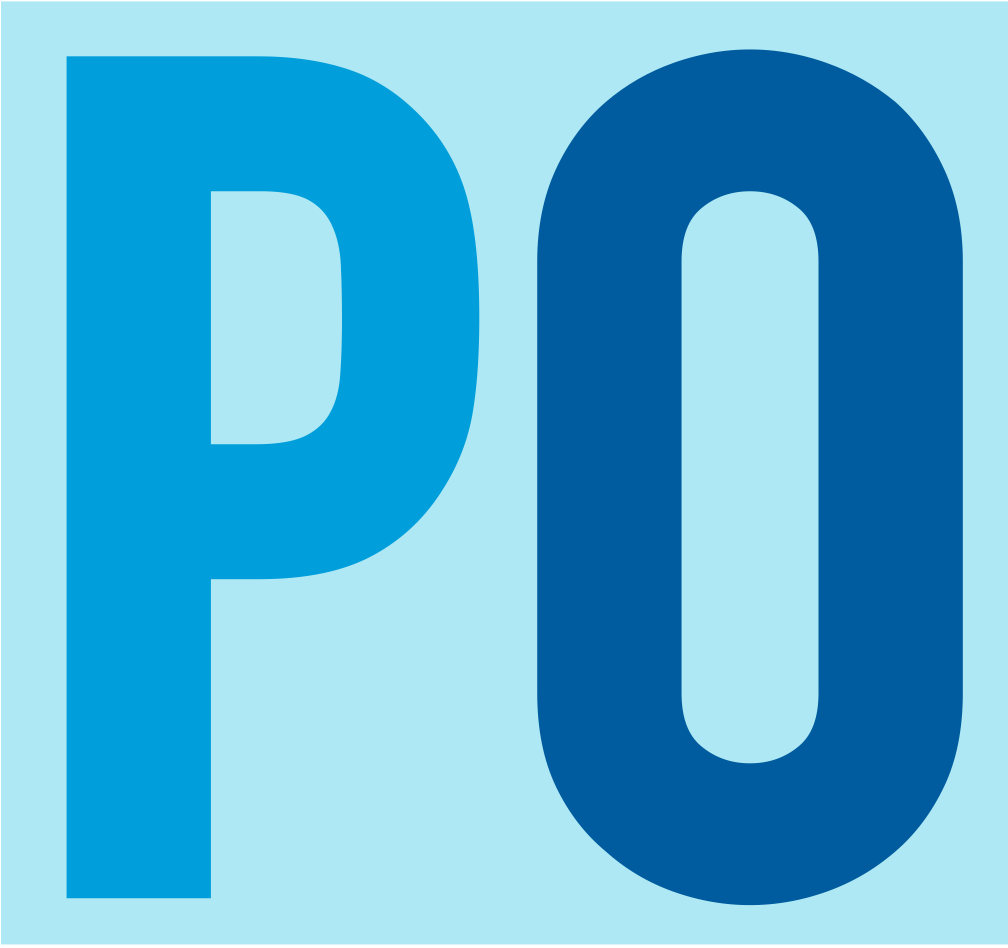}};
\draw[line width=0.75, color=red] (1, -3) circle (\l); 

\draw [dotted, line width = 1] (1, -2.7) -- (3, -2);
\draw [dotted, line width = 1] (1, -3.3) -- (3, -4);

\pgfmathsetmacro{\n}{4}
\pgfmathsetmacro{\dh}{2 / \n}
\draw (3,-4) -- (5,-4) -- (5,-2) -- (3,-2) -- cycle;
\foreach \i in {1,...,\n}{
\pgfmathsetmacro{\h}{\i * \dh}
\draw (3,\h-4) -- (5,\h-4);
\pgfmathsetmacro{\hcent}{\h - (\dh / 2)}
\draw[fill=gray] (3.15, \hcent-4.1) rectangle +(0.2, 0.2);  
\pgfmathsetmacro{\idx}{int(5 - \i)}
\node at (4.18, \hcent-4) {\fontsize{4.5}{2}\selectfont Platform Inst. \idx};
}

\draw [dotted, line width = 1] (5, -2.75) -- (6, -2);
\draw [dotted, line width = 1] (5, -2.75) -- (6, -4);

\draw (6,-4) -- (8,-4) -- (8,-2) -- (6,-2) -- cycle;
\node at (7, 1.8-4) {\fontsize{4.5}{2}\selectfont Platform Instance $k$};
\draw (6, 1.65-4) -- (8, 1.65-4);

\foreach \i in {0, 1}{
    \foreach \j in {0, 1}{
        \pgfmathsetmacro{\dx}{\i * 0.75}
        \pgfmathsetmacro{\dy}{\j * 0.75}
        \draw (6 + \dx + 0.375, -2 - \dy - 0.55) rectangle +(0.5, -0.5);
        \pgfmathsetmacro{\idx}{int(2 * \j + \i + 1)}
        \node at (6.65 + \dx, -3.15 - \dy) {\fontsize{4}{2}\selectfont {Plugin \idx}};
    }
}

\draw (0,-7) -- (2,-7) -- (2,-5) -- (0,-5) -- cycle;
\draw[<->, line width = 1] (1,-5) -- (1,-4);
\node at (1, -5.3) {\fontsize{7}{2}\selectfont Meta-Platform};
\node[anchor=center] at (1, -6.2) {\includegraphics[width=1cm, height=1cm]{tikz_assets/server.png}};

\node[anchor = center, minimum width = 5cm, minimum height = 2cm] (mp) at (5.5, -6) {};
\draw (3, -5) -- (8, -5) -- (8, -7) -- (3, -7) -- cycle;
\draw[<->, line width = 1] (4, -5) -- (4, -4);
\draw[<->, line width = 1] (7, -5) -- (7, -4);
\node at (5.5, -5.3) {\fontsize{7}{2}\selectfont Hosting Options};

\node[anchor=center, yshift=-2mm] at ($(mp.west)!0.2!(mp.east)$) {\includegraphics[width=1cm, height=1cm]{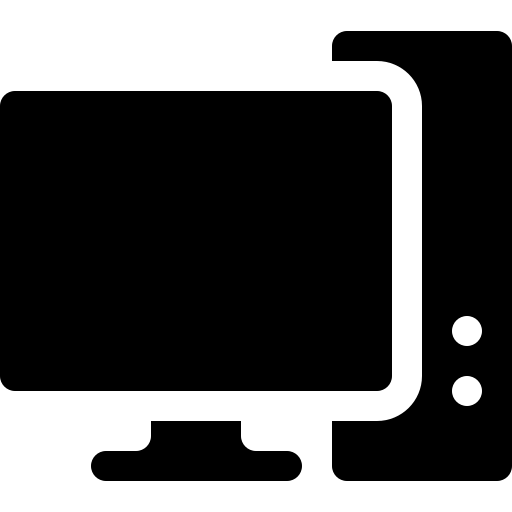}};

\node[anchor=center, yshift=-2mm] at (mp) {\includegraphics[width=1cm, height=1cm]{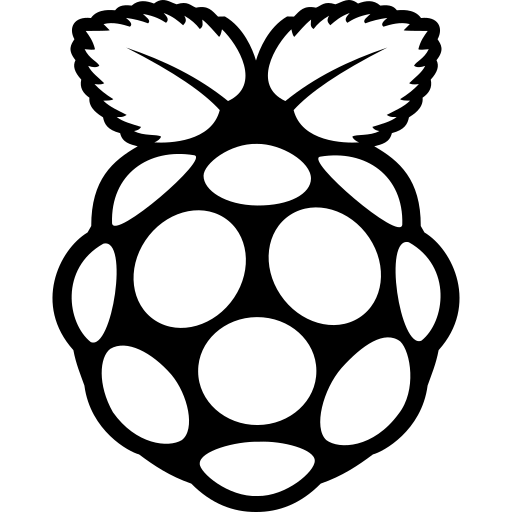}};

\node[anchor=center, yshift=-2mm] at ($(mp.east)!0.2!(mp.west)$) {\includegraphics[width=1cm, height=1cm]{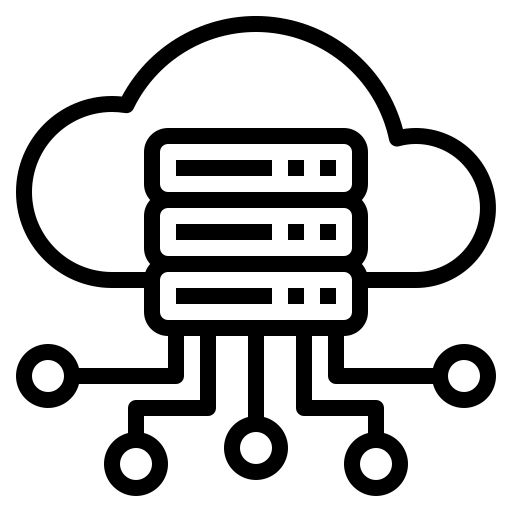}};


\end{tikzpicture}

%% file: tikz_plots/pluginReg.tikz
\begin{tikzpicture}
    \node[cloud, draw, fill=white, aspect=3, align = center, text width = 5cm, scale = 0.8] (mp) at (0, 8) {\Large{Meta Platform} \\ \normalsize{$\textit{addr} \gets \textit{metaPlatform.address}$}};

    \node[rectangle, draw, dotted, line width=1, minimum width=3cm, minimum height=2cm] (cl) at (8,8) {};

    \node[fill = white] at (9, 7) {Client};

    \node[rectangle, draw, fill=gray!30, minimum width=2.5cm, minimum height=1.5cm, text=black] (pl) at (8,8) {Plugin UI};


    \draw[->, line width = 2] (mp.north east) to [out=45,in=135, looseness = 1] node [pos=0.5, above, text width = 4cm] {1.) Download bundling tool} (cl.north west);

    \draw[->, line width = 2] (cl.north) to [out=45,in=45, looseness = 3] node [pos=0.3, above, text width = 4cm] {2.) \textit{inj} $\gets$ \textit{bundle(plugin)}} (cl.east);

    \draw[->, line width = 2] (cl.south west) to [out=225,in=-45, looseness = 1] node [pos=0.5, below, text width = 3cm] {3.) \textbf{post}(\textit{inj, addr})} (mp.south east);

    \draw[->, line width = 2] (mp.north) to [out=135,in=135, looseness = 1.5] node [pos=0.3, above, text width = 4cm] {4.) \textit{acceptPlugin}(\textit{inj})} (mp.west);
    
\end{tikzpicture}

%% file: tikz_plots/pluginArch.tikz
\usetikzlibrary{arrows.meta, positioning, shapes, calc}

\begin{tikzpicture}[
  font=\sffamily,
  every node/.style={align=center},
  serverblock/.style={draw, thick, minimum width=4.5cm, minimum height=5cm, fill=gray!10, rounded corners=5pt},
  plugin/.style={draw, thick, minimum width=2cm, minimum height=1cm, fill=white},
  host/.style={rectangle, draw, thick, minimum width=3cm, minimum height=1.7cm},
  client/.style={rectangle, draw, thick, minimum width=4cm, minimum height=2cm, fill=blue!20},
  remote/.style={rectangle, draw, thick, minimum width=2.5cm, minimum height=1cm, fill=gray!40},
  arrow/.style={-Stealth, thick},
  node distance=1cm and 1cm
]

\node[serverblock] (server) {};
\node at (server.north) [yshift=-0.4cm] {\textbf{Meta-Platform}};

\node[plugin] (pluginA) at ([xshift=-1.1cm, yshift=1.2cm]server.center) {Plugin 1};
\node[plugin] (pluginB) at ([xshift=1.1cm, yshift=1.2cm]server.center) {Plugin 2};
\node[plugin] (pluginC) at ([xshift=-1.1cm, yshift=0cm]server.center) {Plugin 3};
\node[plugin] (pluginD) at ([xshift=1.1cm, yshift=0cm]server.center) {Plugin 4};

\node at ([yshift=-0.8cm]server.center) {$\vdots$};

\node[plugin] (pluginE) at ([xshift=-1.1cm, yshift=-1.8cm]server.center) {Plugin $i-1$};
\node[plugin] (pluginF) at ([xshift=1.1cm, yshift=-1.8cm]server.center) {Plugin $i$};

\filldraw[fill=green!70!black] ([xshift=1.8cm,yshift=-0.9cm]server.center) circle (5pt);

\node[host, left=6cm of pluginB, yshift=0.75cm, label={[yshift=-0.5cm]north:Client 1}] (host1) {};
\node[host, left=6cm of pluginF, yshift=-0.75cm, label={[yshift=-0.5cm]north:Client 2}] (host2) {};
\node[host, right=6cm of pluginA, yshift=0.75cm, label={[yshift=-0.5cm]north:Client $j-1$}] (host3) {};
\node[host, right=6cm of pluginE, yshift=-0.75cm, label={[yshift=-0.5cm]north:Client $j$}] (host4) {};

\node[remote, yshift=-0.15cm] at (host1) {\textit{Remote Code}};
\node[remote, yshift=-0.15cm] at (host2) {\textit{Remote Code}};
\node[remote, yshift=-0.15cm] at (host3) {\textit{Remote Code}};
\node[remote, yshift=-0.15cm] at (host4) {\textit{Remote Code}};

\draw[arrow] (host1.east) to[out=45, in=135] node[pos=0.5, below] {\textbf{fetch}} node[pos=0.5, above] {\footnotesize\texttt{PLUGIN/remoteEntry.js}} ($(server.west)!0.75!(server.north west)$);
\draw[arrow, dotted] ($(server.west)!0.75!(server.north west)$) to[out=225, in=-45] node[pos=0.5, below] {\textit{load dynamically}} (host1.east);

\draw[arrow, dotted] (pluginC.west) -- (host2.east);
\draw[arrow, dotted] (pluginE) -- (host2.east);

\draw[arrow, dotted] (pluginB.east) -- (host3.west);
\draw[arrow, dotted] (pluginD.east) -- (host3.west);

\draw[arrow, dotted] (pluginB.east) -- (host4.west);
\draw[arrow, dotted] (pluginD.east) -- (host4.west);

\node[draw, rectangle, above=2cm of server, minimum width=2.5cm, minimum height=1.5cm] (platform1) {Platform Instance};
\draw[Stealth-Stealth, thick] (platform1.west) to node[pos=0.5, above, sloped] {Distributes datagrams} (host1.north);
\draw[Stealth-Stealth, thick] (platform1.east) to node[pos=0.5, above, sloped] {Distributes datagrams} (host3.north);
\draw[Stealth-, thick] (platform1.south) to node[pos=0.5, right] {Instantiates} (server.north);

\node[draw, rectangle, below=2cm of server, minimum width=2.5cm, minimum height=1.5cm] (platform2) {Platform Instance};
\draw[Stealth-Stealth, thick] (platform2.west) to node[pos=0.5, below, sloped] {Distributes datagrams} (host2.south);
\draw[Stealth-Stealth, thick] (platform2.east) to node[pos=0.5, below, sloped] {Distributes datagrams} (host4.south);
\draw[Stealth-, thick] (platform2.north) to node[pos=0.5, right] {Instantiates} (server.south);


\node[anchor=north west] at (current bounding box.north west) {
  \begin{tikzpicture}[font=\small, baseline]
    \draw[thick, -Stealth] (0,0) -- +(1.2,0) node[right] {Static connection};
    \draw[thick, dotted, -Stealth] (0,-0.5) -- +(1.2,0) node[right] {Dynamic injection};
    \draw[thick, Stealth-Stealth] (0,-1.0) -- +(1.2,0) node[right] {Bidirectional};
  \end{tikzpicture}
};

\end{tikzpicture}